\documentclass[useAMS,usenatbib,article]{mn2e}

\usepackage{longtable}
\usepackage[]{graphicx}
\usepackage{amsmath}
\usepackage{natbib}
\usepackage{amsmath}
\title[The role of dry mergers in the $M_{\rm BH} - \sigma$ diagram]{Overmassive black holes in the $M_{\rm BH} - \sigma$ diagram do not belong to over (dry) merged galaxies}

\author[G.~A.~D. Savorgnan \& A.~W. Graham]
{\parbox{\textwidth}{
Giulia~A.~D. Savorgnan$^{1}$\thanks{E-mail: \texttt{gsavorgn@astro.swin.edu.au}},
Alister W. Graham$^{1}$}\vspace{0.4cm}\\
\parbox{\textwidth}{
$^{1}$Centre for Astrophysics and Supercomputing, Swinburne University of Technology, Hawthorn, Victoria 3122, Australia.\\}}

\pagerange{\pageref{firstpage}--\pageref{lastpage}} \pubyear{2014}

\begin{document}

\maketitle

\label{firstpage}

\begin{abstract}
Semi-analytical models in a $\Lambda$CDM cosmology have predicted the presence of outlying, ``overmassive'' black holes
at the high-mass end of the (black hole mass -- galaxy velocity dispersion) $M_{\rm BH} - \sigma$ diagram
(which we update here with a sample of 89 galaxies).
They are a consequence of having experienced more dry mergers 
-- thought not to increase a galaxy's velocity dispersion --
than the ``main-sequence'' population. 
Wet mergers and gas-rich processes, on the other hand, preserve the main correlation. 
Due to the scouring action of binary supermassive black holes, 
the extent of these dry mergers (since the last significant wet merger) can be traced 
by the ratio between the central stellar mass deficit and the black hole mass ($M_{\rm def,*} / M_{\rm BH}$).
However, in a sample of 23 galaxies with partially depleted cores,
including central cluster galaxies,
we show that the ``overmassive'' black holes 
are actually hosted by galaxies that appear to have undergone the lowest degree of such merging.
In addition, the rotational kinematics of 37 galaxies in the $M_{\rm BH} - \sigma$ diagram 
reveals that fast and slow rotators are not significantly offset from each other,
also contrary to what is expected if these two populations were the product of wet and dry mergers respectively.
The observations are thus not in accordance with model predictions and further investigation is required. 
\end{abstract}


\section{Introduction}
\label{sec:intro}
Our growing awareness of substructures and the actual relations within various black hole mass ($M_{\rm BH}$) scaling diagrams 
is important because it provides us with clues into the joint evolution of black hole and host spheroid. 
For example, \cite{graham2012bent}, \cite{grahamscott2013} and \cite{scott2013}
have shown that the bent $M_{\rm BH} - M_{\rm sph,dyn}$ (spheroid dynamical mass),
$M_{\rm BH} - L_{\rm sph}$ (spheroid luminosity) and
$M_{\rm BH} - M_{\rm sph,*}$ (spheroid stellar mass) relations 
reveal that black holes grow roughly quadratically with their host spheroid until the onset of dry merging,
as signalled by the presence of partially depleted galaxy cores and a linear 
scaling at the high-mass end of these diagrams.
The clever many-merger model of \citet{peng2007}, \cite{hirschmann2010} and \citet{jahnkemaccio2011} was
therefore ruled out because it required convergence along a distribution in the
$M_{\rm BH} - M_{\rm sph,*}$ diagram with a slope of unity, rather than the observed
buildup (to higher masses) along the quadratic relation.\\
In addition, the demographics in the $M_{\rm BH} - \sigma$ (stellar velocity dispersion) 
diagram \citep{ferraresemerritt2000,gebhardt2000} have disclosed a tendency for barred galaxies to be offset, 
to higher velocity dispersions, than non-barred galaxies \citep{graham2007baas,graham2008FP,graham2008bar,hu2008,grahamli2009}.  
This may well be due to the elevated kinematics associated with bars 
(e.g. \citealt{graham2011,brown2013,hartmann2013}).  
Speculation as to the role played by secular evolution and the possibility of ``anaemic'' black holes in pseudo-bulges 
(e.g. \citealt{graham2008FP,hu2008}) does however still remain an intriguing possibility \citep{kormendybender2011},
although their current lack of an offset about the bent $M_{\rm BH} - M_{\rm sph,*}$ relation 
\citep{grahamscott2013} argues against this.  \\
An interesting suggestion for the presence of additional sub-structure in the $M_{\rm BH} - \sigma$ diagram 
has recently been offered by \cite{volontericiotti2013}, 
who investigated why central cluster galaxies tend to be outliers, 
hosting black holes that appear to be ``overmassive'' compared to 
expectations from their velocity dispersion.
On theoretical grounds it is well known that -- 
as a consequence of the virial theorem and the conservation of the total energy -- 
the mass, luminosity and size of a spheroidal galaxy 
increases more readily than its velocity dispersion when a galaxy undergoes  
(parabolic)\footnote{In a parabolic dissipationless merger between
two spheroidal galaxies, the virial velocity dispersion of the merger product cannot be larger than
the maximum velocity dispersion of the progenitors. 
Therefore, when we say that, after such a merger, 
a galaxy experiences a growth of its black hole mass at a fixed velocity dispersion,
we are referring to the progenitor galaxy with the highest velocity dispersion.} 
dissipationless mergers with other spheroidal galaxies 
(e.g. \citealt{ciottivanalbada2001,nipoti2003,ciotti2007,naab2009}).
In this scenario, the supermassive black hole grows through black hole binary merger events, 
while the galaxy velocity dispersion remains unaffected, 
moving the black hole/galaxy pair upward in the $M_{\rm BH} - \sigma$ 
diagram.
Using a combination of analytical and semi-analytical models, \cite{volontericiotti2013} show that 
central cluster galaxies can naturally become outliers in the $M_{\rm BH} - \sigma$ diagram 
because they experience more mergers with spheroidal systems than any other galaxy 
and because these mergers are preferentially gas-poor. \\
Here we test this interesting idea with the latest observational data. 
In so doing, we update the $M_{\rm BH} - \sigma$ diagram to include 89 galaxies now reported to have
directly measured black hole masses.

\section{Rationale}
The high-mass end of the $M_{\rm BH} - \sigma$ diagram, 
where a few ``overmassive'' outliers have now been reported to exist, 
is mainly populated by core-S\'ersic galaxies \citep{graham2003coresersicmodel,trujillo2004coresersicmodel}, 
i.e. galaxies (or bulges) with partially depleted cores
relative to their outer S\'ersic light profile.
While these galaxies are also ``core galaxies'', as given by the Nuker definition \citep{lauer2007lumell},
it should be noted that $\sim$$20\%$ of ``core galaxies'' are not core-S\'ersic galaxies 
(\citealt{dullograham2014cores}, their Appendix A.2), i.e. do not have depleted cores. 
Such S\'ersic galaxies have no central deficit of stars. 
It has long been hypothesized that the presence of a partially depleted core indicates that the host galaxy
has experienced one or more ``dry'' major mergers \citep{begelman1980}. 
During such dissipationless mergers, the progenitor supermassive black holes are expected to sink towards
the centre of the remnant, 
form a bound pair and release their binding energy to the surrounding stars 
(\citealt{milosavljevicmerritt2001,merritt2013CQG} and references therein).
Indeed, the latest high-resolution observations 
(e.g. \citealt{sillanpaa1988,komossa2003,maness2004,rodriguez2006,dotti2009,burke2011,fabbiano2011,ju2013,liu2014}) 
are providing us with compelling evidence of tight black hole binary systems.
The evacuation of stars takes place within the so-called ``loss-cone'' 
of the black hole binary 
and has the effect of lowering the galaxy's central stellar density
(e.g. \citealt{merritt2006RPP}, his Figure 5; \citealt{dotti2012,colpi2014}). 
Upon analyzing the central stellar kinematics of a sample of core galaxies,
\cite{thomas2014} concluded that the homology of the distribution of the orbits matches 
the predictions from black hole binary theoretical models, 
and argued that the small values of central rotation velocities favor a sequence of several minor mergers
rather than a few equal-mass mergers. 
Subsequent to the dry merging events, AGN feedback likely prevents further star formation 
in the spheroids of the core-S\'ersic galaxies (e.g. \citealt{ciotti2010}, and references therein).      
High-accuracy $N$-body simulations \citep{merritt2006}
have shown that, after $\mathcal{N}$ (equivalent) major mergers, the magnitude of the stellar mass deficit $M_{\rm def,*}$ 
scales as $\mathcal{N}$ times the final mass of the relic black hole ($M_{\rm def,*} \approx 0.5 \mathcal{N} M_{\rm BH}$).
This result has been used to make inferences about the galaxy merger history 
(e.g. \citealt{graham2004,ferrarese2006acsvcs,hyde2008,dullograham2014cores}). \\
If one assumes that the ``overmassive'' black holes 
belong to galaxies that have undergone a larger number of dry mergers 
compared to galaxies that obey the observed $M_{\rm BH} - \sigma$ correlation \citep{mcconnellma2013,grahamscott2013},  
it is a natural expectation that 
these $M_{\rm BH} - \sigma$ outliers may also display a higher $M_{\rm def,*} / M_{\rm BH}$ ratio 
when compared to the ``main-sequence'' population. 
This argument motivates our first test. \\
A second test can be built by looking at the kinematics of the objects that populate the $M_{\rm BH} - \sigma$ diagram.
A galaxy's velocity dispersion remains unaffected only in the case of a dissipationless merger 
(with another spheroidal galaxy), 
whereas it accordingly increases after a dissipational (gas-rich) merger,
preserving the $M_{\rm BH} - \sigma$ correlation \citep{volontericiotti2013}.
Wet and dry mergers may produce remnants with different kinematical structures,
classified as fast (disc) and slow rotators, respectively (e.g. \citealt{emsellem2008fastslow} and references therein).
Therefore, an instinctive question is whether the populations of slow and fast rotators 
are significantly offset from each other in the $M_{\rm BH} - \sigma$ diagram. 
This will be our second test.

\section{Data}
\label{sec:data}
\begin{figure}
\begin{center}
\includegraphics[width=\columnwidth, trim=0 30 0 0]{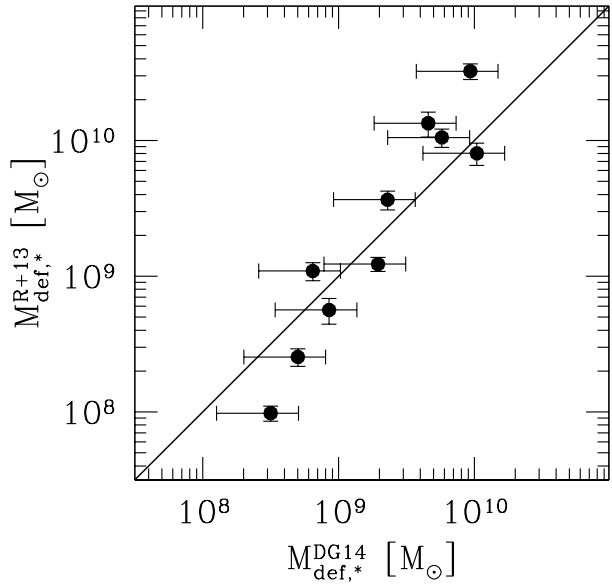}
\end{center}
\caption{Comparison between the central stellar mass deficits estimated by \citeauthor{rusli2013} 
(\citeyear{rusli2013}, R+13) and \citeauthor{dullograham2014cores} (\citeyear{dullograham2014cores}, DG14)
for ten galaxies in common. 
The black solid line shows the 1:1 relation. 
As noted in the text, the small ``apparent'' systematic difference is actually due to random causes.}
\label{fig:cfr}
\end{figure}

Our galaxy sample (see Table \ref{tab:data}) consists of 89 objects for which a dynamical detection of the black hole mass
and a measure of the stellar velocity dispersion have been reported in the literature.
We include in our sample all the 78 objects presented in the catalog of \cite{grahamscott2013},
plus 10 objects taken from \cite{rusli2013bhmassesDM} and 1 object from \cite{greenhill2003}.
Partially depleted cores have been identified according to the same 
criteria used by \cite{grahamscott2013}.
When no high-resolution image analysis was available from the literature, 
we inferred the presence of a partially depleted core based on the stellar velocity dispersion, $\sigma$: 
a galaxy is classified as core-S\'ersic if $\sigma > 270 \rm~km~s^{-1}$, or as 
S\'ersic if $\sigma \leq 166 \rm~km~s^{-1}$. 
This resulted in us assigning cores to just 6 galaxies, 
none of which were used in the following mass deficit analysis.
We employ a $5\%$ uncertainty on $\sigma$ in our regression analysis. \\
A kinematical classification (slow/fast rotator) is available for 34 of our 89 galaxies from the ATLAS$^{\rm 3D}$ survey \citep{atlas3dIII}
and for 3 additional 
galaxies\footnote{NGC 1316, NGC 1374 and NGC 1399.} from \cite{scott2014}. 
It is however beyond the scope of this paper to derive slow/fast rotator classifications for the remaining galaxies. \\
All galaxies are categorised as barred/unbarred objects
according to the classification reported by \cite{grahamscott2013},
with the following updates.
An isophotal analysis and unsharp masking of {\emph Spitzer}/IRAC $3.6~\rm \mu m$ images 
(Savorgnan et al. \emph{in preparation}) has revealed the presence of a bar in the galaxies 
NGC 0224 (in agreement with \citealt{athanassoulabeaton2006m31,beaton2007m31,morrison2011m31}),
NGC 2974 (confirming the suggestion of \citealt{jeong2007}), 
NGC 3031 (see also \citealt{elmegreen1995m81,gutierrez2011,erwindebattista2013}),
NGC 3245 (see also \citealt{laurikainen2010,gutierrez2011}),
NGC 3998 (as already noted by \citealt{gutierrez2011}), 
NGC 4026, 
NGC 4388 and
NGC 4736 (see also \citealt{moellenhoff1995}).\\
Although the fast rotator galaxy NGC 1316 has been frequently classified in the literaure as an elliptical merger remnant,
\cite{grahamscott2013} identified this object as a barred lenticular galaxy. 
\cite{donofrio2001} found that a single-component model cannot provide a good description of the light profile
of this galaxy and
\cite{desouza2004} fit NGC 1316 with a bulge + exponential disc model.
\cite{sani2011} adopted a three-component model, featuring a bulge, an exponential disc and a central Gaussian (attributed
to non-stellar nuclear emission).
Upon an analysis of the two-dimensional velocity field obtained from the kinematics of planetary nebulae,
\cite{mcneilmoylan2012} claimed that NGC 1316 represents a transition phase from a major-merger event 
to a bulge-dominated galaxy like the Sombrero galaxy (M104).
We find evidence for the presence of a bar 
in NGC 1316 from an isophotal analysis and unsharp masking 
of its {\emph Spitzer}/IRAC $3.6~\rm \mu m$ image (Savorgnan et al. \emph{in preparation}), 
but we exclude it for now to avoid any controversy.
\\
Central stellar mass deficits (with individual uncertainties) have been estimated for 23 core-S\'ersic galaxies 
-- with directly measured black hole masses -- by \cite{rusli2013}.
Briefly, they fit the surface brightness profiles of these galaxies with a core-S\'ersic model 
and computed the light deficit as the difference between the luminosity of the S\'ersic 
component of the best-fitting core-S\'ersic model and the luminosity of the core-S\'ersic model itself.
Light deficits were then converted into stellar mass deficits through 
dynamically-determined, individual stellar mass-to-light ratios.
\cite{rusli2013} used galaxy distances sligthly different from those adopted in this work 
(see Table \ref{tab:data}), therefore we adjusted their stellar mass deficits (and uncertainties) 
accordingly\footnote{Mass deficits and their uncertainties from \cite{rusli2013}
were corrected by a factor of $(D/D_{\rm prev})$.
Mass deficits from \cite{dullograham2014cores} were corrected by a factor of $(D/D_{\rm prev})^2$.
Here, $D$ are the galaxy distances
adopted in this work and $D_{\rm prev}$ are the galaxy distances used in the original works.}.
Among the 23 core-S\'ersic galaxies whose stellar mass deficits have been computed by \cite{rusli2013},
10 were also analyzed by \cite{dullograham2014cores}. 
\cite{dullograham2014cores} measured light deficits with a method similar to that
employed by \cite{rusli2013}, but they converted light deficits into stellar mass deficits using stellar mass-to-light
ratios derived from $V-I$ colours together with the color-age-metallicity diagram \citep{grahamspitler2009}. 
Their stellar mass deficits are accurate to $60\%$ (Dullo, private communication) and were rescaled according 
to the galaxy distances adopted here.
In Figure \ref{fig:cfr} we compare these 10 common mass deficit estimates.
The agreement is remarkably good, although
a slight deviation from the 1:1 line can be noticed for the galaxies with the lowest or highest mass deficits,
for which $M_{\rm def,*}$ reported by \cite{dullograham2014cores} is %systematically 
larger or smaller than \cite{rusli2013}, respectively.
We checked and found that this effect actually depends in a random, i.e. non-systematic, way 
on the different choices to estimate the stellar mass-to-light ratios
and/or their different galaxy data and modelling.
We return to this point in the next Section.
For these individual 10 galaxies 
we compute a weighted arithmetic mean of their two available stellar mass deficits. \\

\begin{table*}
\begin{center}
\caption{{\bf Galaxy sample.}
\emph{Column (1)}: Galaxy names; for the 18 galaxies marked with a *, the black hole masses were estimated including in the modelling the effects of dark matter.
\emph{Column (2)}: Distances. 
\emph{Column (3)}: Black hole masses; for the 10 measurements taken from \citet{rusli2013bhmassesDM}, we report in parenthesis also the measurements obtained
without including in the modelling the effects of dark matter. 
\emph{Column (4)}: Stellar velocity dispersions. 
\emph{Column (5)}: References of black hole mass and velocity dispersion measurements reported here (G+03 = \citealt{greenhill2003}, R+13 = \citealt{rusli2013bhmassesDM}, GS13 = \citealt{grahamscott2013}). 
\emph{Column (6)}: Presence of a partially depleted core. 
The question mark is used when the classification has come from the velocity dispersion criteria mentioned in Section \ref{sec:data}.
\emph{Column (7)}: Presence of a bar. 
\emph{Column (8)}: Central stellar mass deficits as measured by \citet{rusli2013}. 
For 7 galaxies we reconstructed the ``no-dark-matter'' values (see Section \ref{sec:mbh}), which are reported in parenthesis.
\emph{Column (9)}: Central stellar mass deficits as measured by \citet{dullograham2014cores}. 
\emph{Column (10)}: Kinematical classification (fast/slow rotator).}
\begin{tabular}{llllllllll}
\hline
{\bf Galaxy}      &  {\bf Dist} & $\mathbf M_{\rm \bf BH}$  &   $\boldsymbol \sigma$  &  {\bf Ref.} &  {\bf Core}  &  {\bf Bar}   &  $\mathbf M_{\rm \bf def,*}^{\rm \bf R+13}$  & $\mathbf M_{\rm \bf def,*}^{\rm \bf DG13}$  &  {\bf Kinematics}  \\ 
                  &  $\rm Mpc$  & $[10^8 \rm~M_{\odot}]$    & $[\rm km~s^{-1}]$       &             &              &              &  $[10^8 \rm~M_{\odot}]$                      &  $[10^8 \rm~M_{\odot}]$ &   \\
(1) & (2)  & (3) & (4) & (5)  & (6)  & (7)  & (8) & (9) & (10)  \\
\hline
    A1836-BCG                       &           $       158.0           $           &           $          39       ^{+4}       _{-5}           $       \quad           $                       $           &           $         309      ^{+15}      _{-15}           $           &        GS13           &        yes?           &   	no		 &	     $  					   $	   \quad	   $			   $	       &	   $						   $	       &	   -	      \\
    A3565-BCG                       &           $        40.7           $           &           $          11       ^{+2}       _{-2}           $       \quad           $                       $           &           $         335      ^{+17}      _{-17}           $           &        GS13           &        yes?           &   	no		 &	     $  					   $	   \quad	   $			   $	       &	   $						   $	       &	   -	      \\
    Circinus                        &           $         4.0           $           &           $       0.017   ^{+0.004}   _{-0.003}           $       \quad           $                       $           &           $         158       ^{+8}       _{-8}           $           &        G+03           &         no?           &   	no		 &	     $  					   $	   \quad	   $			   $	       &	   $						   $	       &	   -	      \\
     CygnusA                        &           $       232.0           $           &           $          25       ^{+7}       _{-7}           $       \quad           $                       $           &           $         270      ^{+13}      _{-13}           $           &        GS13           &        yes?           &   	no		 &	     $  					   $	   \quad	   $			   $	       &	   $						   $	       &	   -	      \\
      IC 1459                       &           $        28.4           $           &           $          24      ^{+10}      _{-10}           $       \quad           $                       $           &           $         306      ^{+15}      _{-15}           $           &        GS13           &         yes           &   	no		 &	     $  	16	 ^{+7}       _{-7}	   $	   \quad	   $			   $	       &	   $						   $	       &	   -	      \\
      IC 2560                       &           $        40.7           $           &           $       0.044   ^{+0.044}   _{-0.022}           $       \quad           $                       $           &           $         144       ^{+7}       _{-7}           $           &        GS13           &         no?           &          yes		 &	     $  					   $	   \quad	   $			   $	       &	   $						   $	       &	   -	      \\
         M32                        &           $         0.8           $           &           $       0.024   ^{+0.005}   _{-0.005}           $       \quad           $                       $           &           $          55       ^{+3}       _{-3}           $           &        GS13           &          no           &   	no		 &	     $  					   $	   \quad	   $			   $	       &	   $						   $	       &	   -	      \\
    Milky Way                       &           $         0.008         $           &           $       0.043   ^{+0.004}   _{-0.004}           $       \quad           $                       $           &           $         100       ^{+5}       _{-5}           $           &        GS13           &          no           &          yes		 &	     $  					   $	   \quad	   $			   $	       &	   $						   $	       &	   -	      \\
     NGC 0224	  		    &		$	  0.7		$	    &		$	  1.4	  ^{+0.9}     _{-0.3}		$	\quad		$			$	    &		$	  170	    ^{+8}	_{-8}		$	    &	     GS13	    &	       no	    &          yes		 &	     $  					   $	   \quad	   $			   $	       &	   $						   $	       &	   -	      \\
     NGC 0253	  		    &		$	  3.5		$	    &		$	  0.1	  ^{+0.1}    _{-0.05}		$	\quad		$			$	    &		$	  109	    ^{+5}	_{-5}		$	    &	     GS13	    &	       no	    &          yes		 &	     $  					   $	   \quad	   $			   $	       &	   $						   $	       &	   -	      \\
     NGC 0524	  		    &		$	 23.3		$	    &		$	  8.3	  ^{+2.7}     _{-1.3}		$	\quad		$			$	    &		$	  253	   ^{+13}      _{-13}		$	    &	     GS13	    &	      yes	    &   	no		 &	     $  					   $	   \quad	   $			   $	       &	   $						   $	       &	FAST	      \\
     NGC 0821	  		    &		$	 23.4		$	    &		$	 0.39	 ^{+0.26}    _{-0.09}		$	\quad		$			$	    &		$	  200	   ^{+10}      _{-10}		$	    &	     GS13	    &	       no	    &   	no		 &	     $  					   $	   \quad	   $			   $	       &	   $						   $	       &	FAST	      \\
     NGC 1023	  		    &		$	 11.1		$	    &		$	 0.42	 ^{+0.04}    _{-0.04}		$	\quad		$			$	    &		$	  204	   ^{+10}      _{-10}		$	    &	     GS13	    &	       no	    &          yes		 &	     $  					   $	   \quad	   $			   $	       &	   $						   $	       &	FAST	      \\
     NGC 1068	  		    &		$	 15.2		$	    &		$	0.084	^{+0.003}   _{-0.003}		$	\quad		$			$	    &		$	  165	    ^{+8}	_{-8}		$	    &	     GS13	    &	       no	    &          yes		 &	     $  					   $	   \quad	   $			   $	       &	   $						   $	       &	   -	      \\
     NGC 1194	  		    &		$	 53.9		$	    &		$	 0.66	 ^{+0.03}    _{-0.03}		$	\quad		$			$	    &		$	  148	    ^{+7}	_{-7}		$	    &	     GS13	    &	      no?	    &   	no		 &	     $  					   $	   \quad	   $			   $	       &	   $						   $	       &	   -	      \\
     NGC 1300	  		    &		$	 20.7		$	    &		$	 0.73	 ^{+0.69}    _{-0.35}		$	\quad		$			$	    &		$	  229	   ^{+11}      _{-11}		$	    &	     GS13	    &	       no	    &          yes		 &	     $  					   $	   \quad	   $			   $	       &	   $						   $	       &	   -	      \\
     NGC 1316	  		    &		$	 18.6		$	    &		$	  1.5	 ^{+0.75}     _{-0.8}		$	\quad		$			$	    &		$	  226	   ^{+11}      _{-11}		$	    &	     GS13	    &	       no	    &          ?		 &	     $  					   $	   \quad	   $			   $	       &	   $						   $	       &	FAST	      \\
     NGC 1332	  		    &		$	 22.3		$	    &		$	 14.5	    ^{+2}	_{-2}		$	\quad		$			$	    &		$	  320	   ^{+16}      _{-16}		$	    &	     GS13	    &	       no	    &   	no		 &	     $  					   $	   \quad	   $			   $	       &	   $						   $	       &	   -	      \\
     NGC 1374	  	*	    &		$	 19.2		$	    &		$	  5.8	  ^{+0.5}     _{-0.5}		$	\quad		$	(5.8)		$	    &		$	  167	    ^{+8}	_{-8}		$	    &	     R+13	    &	      no?	    &   	no		 &	     $  					   $	   \quad	   $			   $	       &	   $						   $	       &	FAST	      \\
     NGC 1399	  		    &		$	 19.4		$	    &		$	  4.7	  ^{+0.6}     _{-0.6}		$	\quad		$			$	    &		$	  329	   ^{+16}      _{-16}		$	    &	     GS13	    &	      yes	    &   	no		 &	     $         324	^{+42}      _{-42}	   $	   \quad	   $			   $	       &	   $	      93      ^{+56}	  _{-56}	   $	       &	SLOW	      \\
     NGC 1407	  	*	    &		$	 28.1		$	    &		$	   45	    ^{+4}	_{-9}		$	\quad		$	 (38)		$	    &		$	  276	   ^{+14}      _{-14}		$	    &	     R+13	    &	      yes	    &   	no		 &	     $  	43	 ^{+9}       _{-9}	   $	   \quad	   $	    (70)	   $	       &	   $						   $	       &	   -	      \\
     NGC 1550	  	*	    &		$	 51.6		$	    &		$	   37	    ^{+4}	_{-4}		$	\quad		$	 (37)		$	    &		$	  270	   ^{+13}      _{-13}		$	    &	     R+13	    &	      yes	    &   	no		 &	     $         109	^{+19}      _{-19}	   $	   \quad	   $	   (117)	   $	       &	   $						   $	       &	   -	      \\
     NGC 2273	  		    &		$	 28.5		$	    &		$	0.083	^{+0.004}   _{-0.004}		$	\quad		$			$	    &		$	  145	    ^{+7}	_{-7}		$	    &	     GS13	    &	       no	    &          yes		 &	     $  					   $	   \quad	   $			   $	       &	   $						   $	       &	   -	      \\
     NGC 2549	  		    &		$	 12.3		$	    &		$	 0.14	 ^{+0.02}    _{-0.13}		$	\quad		$			$	    &		$	  144	    ^{+7}	_{-7}		$	    &	     GS13	    &	       no	    &          yes		 &	     $  					   $	   \quad	   $			   $	       &	   $						   $	       &	FAST	      \\
     NGC 2778	  		    &		$	 22.3		$	    &		$	 0.15	 ^{+0.09}     _{-0.1}		$	\quad		$			$	    &		$	  162	    ^{+8}	_{-8}		$	    &	     GS13	    &	       no	    &          yes		 &	     $  					   $	   \quad	   $			   $	       &	   $						   $	       &	FAST	      \\
     NGC 2787	  		    &		$	  7.3		$	    &		$	  0.4	 ^{+0.04}    _{-0.05}		$	\quad		$			$	    &		$	  210	   ^{+10}      _{-10}		$	    &	     GS13	    &	       no	    &          yes		 &	     $  					   $	   \quad	   $			   $	       &	   $						   $	       &	   -	      \\
     NGC 2960	  		    &		$	 81.0		$	    &		$	0.117	^{+0.005}   _{-0.005}		$	\quad		$			$	    &		$	  166	    ^{+8}	_{-8}		$	    &	     GS13	    &	      no?	    &   	no		 &	     $  					   $	   \quad	   $			   $	       &	   $						   $	       &	   -	      \\
     NGC 2974	  		    &		$	 20.9		$	    &		$	  1.7	  ^{+0.2}     _{-0.2}		$	\quad		$			$	    &		$	  227	   ^{+11}      _{-11}		$	    &	     GS13	    &	       no	    &          yes		 &	     $  					   $	   \quad	   $			   $	       &	   $						   $	       &	FAST	      \\
     NGC 3031	  		    &		$	  3.8		$	    &		$	 0.74	 ^{+0.21}    _{-0.11}		$	\quad		$			$	    &		$	  162	    ^{+8}	_{-8}		$	    &	     GS13	    &	       no	    &          yes		 &	     $  					   $	   \quad	   $			   $	       &	   $						   $	       &	   -	      \\
     NGC 3079	  		    &		$	 20.7		$	    &		$	0.024	^{+0.024}   _{-0.012}		$	\quad		$			$	    &		$	60-150                   		$	    &	     GS13	    &	      no?	    &          yes		 &	     $  					   $	   \quad	   $			   $	       &	   $						   $	       &	   -	      \\
     NGC 3091	  	*	    &		$	 51.3		$	    &		$	   36	    ^{+2}	_{-1}		$	\quad		$	(9.7)		$	    &		$	  297	   ^{+15}      _{-15}		$	    &	     R+13	    &	      yes	    &   	no		 &	     $         157	^{+34}      _{-34}	   $	   \quad	   $	   (224)	   $	       &	   $						   $	       &	   -	      \\
     NGC 3115	  		    &		$	  9.4		$	    &		$	  8.8	   ^{+10.0}     _{-2.7}		$	\quad		$			$	    &		$	  252	   ^{+13}      _{-13}		$	    &	     GS13	    &	       no	    &   	no		 &	     $  					   $	   \quad	   $			   $	       &	   $						   $	       &	   -	      \\
     NGC 3227	  		    &		$	 20.3		$	    &		$	 0.14	  ^{+0.10}    _{-0.06}		$	\quad		$			$	    &		$	  133	    ^{+7}	_{-7}		$	    &	     GS13	    &	       no	    &          yes		 &	     $  					   $	   \quad	   $			   $	       &	   $						   $	       &	   -	      \\
     NGC 3245	  		    &		$	 20.3		$	    &		$	    2	  ^{+0.5}     _{-0.5}		$	\quad		$			$	    &		$	  210	   ^{+10}      _{-10}		$	    &	     GS13	    &	       no	    &          yes		 &	     $  					   $	   \quad	   $			   $	       &	   $						   $	       &	FAST	      \\
     NGC 3368	  		    &		$	 10.1		$	    &		$	0.073	^{+0.015}   _{-0.015}		$	\quad		$			$	    &		$	  128	    ^{+6}	_{-6}		$	    &	     GS13	    &	       no	    &          yes		 &	     $  					   $	   \quad	   $			   $	       &	   $						   $	       &	   -	      \\
     NGC 3377	  		    &		$	 10.9		$	    &		$	 0.77	 ^{+0.04}    _{-0.06}		$	\quad		$			$	    &		$	  139	    ^{+7}	_{-7}		$	    &	     GS13	    &	       no	    &   	no		 &	     $  					   $	   \quad	   $			   $	       &	   $						   $	       &	FAST	      \\
     NGC 3379	  		    &		$	 10.3		$	    &		$	    4	    ^{+1}	_{-1}		$	\quad		$			$	    &		$	  209	   ^{+10}      _{-10}		$	    &	     GS13	    &	      yes	    &   	no		 &	     $  	12	 ^{+1}       _{-1}	   $	   \quad	   $			   $	       &	   $	      19      ^{+12}	  _{-12}	   $	       &	FAST	      \\
     NGC 3384	  		    &		$	 11.3		$	    &		$	 0.17	 ^{+0.01}    _{-0.02}		$	\quad		$			$	    &		$	  148	    ^{+7}	_{-7}		$	    &	     GS13	    &	       no	    &          yes		 &	     $  					   $	   \quad	   $			   $	       &	   $						   $	       &	FAST	      \\
     NGC 3393	  		    &		$	 55.2		$	    &		$	 0.34	 ^{+0.02}    _{-0.02}		$	\quad		$			$	    &		$	  197	   ^{+10}      _{-10}		$	    &	     GS13	    &	       no	    &          yes		 &	     $  					   $	   \quad	   $			   $	       &	   $						   $	       &	   -	      \\
     NGC 3414	  		    &		$	 24.5		$	    &		$	  2.4	  ^{+0.3}     _{-0.3}		$	\quad		$			$	    &		$	  237	   ^{+12}      _{-12}		$	    &	     GS13	    &	       no	    &   	no		 &	     $  					   $	   \quad	   $			   $	       &	   $						   $	       &	SLOW	      \\
     NGC 3489	  		    &		$	 11.7		$	    &		$	0.058	^{+0.008}   _{-0.008}		$	\quad		$			$	    &		$	  105	    ^{+5}	_{-5}		$	    &	     GS13	    &	       no	    &          yes		 &	     $  					   $	   \quad	   $			   $	       &	   $						   $	       &	FAST	      \\
     NGC 3585	  		    &		$	 19.5		$	    &		$	  3.1	  ^{+1.4}     _{-0.6}		$	\quad		$			$	    &		$	  206	   ^{+10}      _{-10}		$	    &	     GS13	    &	       no	    &   	no		 &	     $  					   $	   \quad	   $			   $	       &	   $						   $	       &	   -	      \\
     NGC 3607	  		    &		$	 22.2		$	    &		$	  1.3	  ^{+0.5}     _{-0.5}		$	\quad		$			$	    &		$	  224	   ^{+11}      _{-11}		$	    &	     GS13	    &	       no	    &   	no		 &	     $  					   $	   \quad	   $			   $	       &	   $						   $	       &	FAST	      \\
     NGC 3608	  		    &		$	 22.3		$	    &		$	    2	  ^{+1.1}     _{-0.6}		$	\quad		$			$	    &		$	  192	   ^{+10}      _{-10}		$	    &	     GS13	    &	      yes	    &   	no		 &	     $       1.0     ^{+0.1}	_{-0.1} 	$	\quad		$			$	    &		$	    3	    ^{+2}	_{-2}		$	    &	     SLOW	   \\
     NGC 3842	  	*	    &		$	 98.4		$	    &		$	   97	   ^{+30}      _{-26}		$	\quad		$			$	    &		$	  270	   ^{+13}      _{-13}		$	    &	     GS13	    &	      yes	    &   	no		 &	     $  	80	^{+15}      _{-15}	   $	\quad		$			$	    &		$	  104	   ^{+63}      _{-63}		$	    &		-	   \\
\hline
%\multicolumn{10}{|r|}{\emph{The full table is made available online in electronic version.}} \\
\end{tabular}
\label{tab:data} 
\end{center}
\end{table*}

\begin{table*}
\begin{center}
\begin{tabular}{llllllllll}
\hline
{\bf Galaxy}	  &  {\bf Dist} & $\mathbf M_{\rm \bf BH}$  &	$\boldsymbol \sigma$  &  {\bf Ref.} &  {\bf Core}  &  {\bf Bar}   &  $\mathbf M_{\rm \bf def,*}^{\rm \bf R+13}$  & $\mathbf M_{\rm \bf def,*}^{\rm \bf DG13}$  &  {\bf Kinematics}  \\ 
		  &  $\rm Mpc$  & $[10^8 \rm~M_{\odot}]$    & $[\rm km~s^{-1}]$       & 	    &		   &		  &  $[10^8 \rm~M_{\odot}]$			 &  $[10^8 \rm~M_{\odot}]$ &   \\
(1) & (2)  & (3) & (4) & (5)  & (6)  & (7)  & (8) & (9) & (10)  \\
\hline
     NGC 3998	  	*	    &		$	 13.7		$	    &		$	  8.1	    ^{+2}     _{-1.9}		$	\quad		$			$	    &		$	  305	   ^{+15}      _{-15}		$	    &	     GS13	    &	       no	    &	  yes		    &		$					    $	    \quad	    $			    $		&	    $						    $		&	 FAST	       \\
     NGC 4026	  		    &		$	 13.2		$	    &		$	  1.8	  ^{+0.6}     _{-0.3}		$	\quad		$			$	    &		$	  178	    ^{+9}	_{-9}		$	    &	     GS13	    &	       no	    &	  yes		    &		$					    $	    \quad	    $			    $		&	    $						    $		&	 FAST	       \\
     NGC 4151	  		    &		$	 20.0		$	    &		$	 0.65	 ^{+0.07}    _{-0.07}		$	\quad		$			$	    &		$	  156	    ^{+8}	_{-8}		$	    &	     GS13	    &	       no	    &	  yes		    &		$					    $	    \quad	    $			    $		&	    $						    $		&	    -	       \\
     NGC 4258	  		    &		$	  7.2		$	    &		$	 0.39	 ^{+0.01}    _{-0.01}		$	\quad		$			$	    &		$	  134	    ^{+7}	_{-7}		$	    &	     GS13	    &	       no	    &	  yes		    &		$					    $	    \quad	    $			    $		&	    $						    $		&	    -	       \\
     NGC 4261	  		    &		$	 30.8		$	    &		$	    5	    ^{+1}	_{-1}		$	\quad		$			$	    &		$	  309	   ^{+15}      _{-15}		$	    &	     GS13	    &	      yes	    &	   no		    &		$	 89	 ^{+15}      _{-15}	    $	    \quad	    $			    $		&	    $						    $		&	 SLOW	       \\
     NGC 4291	  		    &		$	 25.5		$	    &		$	  3.3	  ^{+0.9}     _{-2.5}		$	\quad		$			$	    &		$	  285	   ^{+14}      _{-14}		$	    &	     GS13	    &	      yes	    &	   no		    &		$	 11	  ^{+2}       _{-2}	    $	    \quad	    $			    $		&	    $		6	^{+4}	    _{-4}	    $		&	    -	       \\
     NGC 4342	  		    &		$	 23.0		$	    &		$	  4.5	  ^{+2.3}     _{-1.5}		$	\quad		$			$	    &		$	  253	   ^{+13}      _{-13}		$	    &	     GS13	    &	       no	    &	   no		    &		$					    $	    \quad	    $			    $		&	    $						    $		&	 FAST	       \\
     NGC 4374	  		    &		$	 17.9		$	    &		$	    9.0	  ^{+0.9}     _{-0.8}		$	\quad		$			$	    &		$	  296	   ^{+15}      _{-15}		$	    &	     GS13	    &	      yes	    &	   no		    &		$	 66	 ^{+11}      _{-11}	    $	    \quad	    $			    $		&	    $						    $		&	 SLOW	       \\
     NGC 4388	  		    &		$	 17.0		$	    &		$	0.075	^{+0.002}   _{-0.002}		$	\quad		$			$	    &		$	  107	    ^{+5}	_{-5}		$	    &	     GS13	    &	      no?	    &	  yes		    &		$					    $	    \quad	    $			    $		&	    $						    $		&	    -	       \\
     NGC 4459	  		    &		$	 15.7		$	    &		$	 0.68	 ^{+0.13}    _{-0.13}		$	\quad		$			$	    &		$	  178	    ^{+9}	_{-9}		$	    &	     GS13	    &	       no	    &	   no		    &		$					    $	    \quad	    $			    $		&	    $						    $		&	 FAST	       \\
     NGC 4472	  	*	    &		$	 17.1		$	    &		$	   25	    ^{+1}	_{-3}		$	\quad		$	 (17)		$	    &		$	  300	   ^{+15}      _{-15}		$	    &	     R+13	    &	      yes	    &	   no		    &		$	 37	  ^{+6}       _{-6}	    $	    \quad	    $	     (55)	    $		&	    $	       23      ^{+14}	   _{-14}	    $		&	 SLOW	       \\
     NGC 4473	  		    &		$	 15.3		$	    &		$	  1.2	  ^{+0.4}     _{-0.9}		$	\quad		$			$	    &		$	  179	    ^{+9}	_{-9}		$	    &	     GS13	    &	       no	    &	   no		    &		$					    $	    \quad	    $			    $		&	    $						    $		&	 FAST	       \\
     NGC 4486	  	*	    &		$	 15.6		$	    &		$	   58.0	  ^{+3.5}     _{-3.5}		$	\quad		$			$	    &		$	  334	   ^{+17}      _{-17}		$	    &	     GS13	    &	      yes	    &	   no		    &		$	612	^{+121}     _{-121}	    $	    \quad	    $			    $		&	    $						    $		&	 SLOW	       \\
    NGC 4486a     		    &		$	 17.0		$	    &		$	 0.13	 ^{+0.08}    _{-0.08}		$	\quad		$			$	    &		$	  110	    ^{+5}	_{-5}		$	    &	     GS13	    &	       no	    &	   no		    &		$					    $	    \quad	    $			    $		&	    $						    $		&	 FAST	       \\
     NGC 4552	  		    &		$	 14.9		$	    &		$	  4.7	  ^{+0.5}     _{-0.5}		$	\quad		$			$	    &		$	  252	   ^{+13}      _{-13}		$	    &	     GS13	    &	      yes	    &	   no		    &		$	  3	  ^{+0}       _{-0}	    $	    \quad	    $			    $		&	    $		5	^{+3}	    _{-3}	    $		&	 SLOW	       \\
     NGC 4564	  		    &		$	 14.6		$	    &		$	  0.60	 ^{+0.03}    _{-0.09}		$	\quad		$			$	    &		$	  157	    ^{+8}	_{-8}		$	    &	     GS13	    &	       no	    &	   no		    &		$					    $	    \quad	    $			    $		&	    $						    $		&	 FAST	       \\
     NGC 4594	  	*	    &		$	  9.5		$	    &		$	  6.4	  ^{+0.4}     _{-0.4}		$	\quad		$			$	    &		$	  297	   ^{+15}      _{-15}		$	    &	     GS13	    &	      yes	    &	   no		    &		$					    $	    \quad	    $			    $		&	    $						    $		&	    -	       \\
     NGC 4596	  		    &		$	 17.0		$	    &		$	 0.79	 ^{+0.38}    _{-0.33}		$	\quad		$			$	    &		$	  149	    ^{+7}	_{-7}		$	    &	     GS13	    &	       no	    &	  yes		    &		$					    $	    \quad	    $			    $		&	    $						    $		&	 FAST	       \\
     NGC 4621	  		    &		$	 17.8		$	    &		$	  3.9	  ^{+0.4}     _{-0.4}		$	\quad		$			$	    &		$	  225	   ^{+11}      _{-11}		$	    &	     GS13	    &	       no	    &	   no		    &		$					    $	    \quad	    $			    $		&	    $						    $		&	 FAST	       \\
     NGC 4649	  	*	    &		$	 16.4		$	    &		$	   47	   ^{+10}      _{-10}		$	\quad		$			$	    &		$	  335	   ^{+17}      _{-17}		$	    &	     GS13	    &	      yes	    &	   no		    &		$	105	 ^{+16}      _{-16}	    $	    \quad	    $			    $		&	    $	       58      ^{+35}	   _{-35}	    $		&	 FAST	       \\
     NGC 4697	  		    &		$	 11.4		$	    &		$	  1.8	  ^{+0.2}     _{-0.1}		$	\quad		$			$	    &		$	  171	    ^{+9}	_{-9}		$	    &	     GS13	    &	       no	    &	   no		    &		$					    $	    \quad	    $			    $		&	    $						    $		&	 FAST	       \\
     NGC 4736	  		    &		$	  4.4		$	    &		$	 0.060	^{+0.014}   _{-0.014}		$	\quad		$			$	    &		$	  104	    ^{+5}	_{-5}		$	    &	     GS13	    &	      no?	    &	  yes		    &		$					    $	    \quad	    $			    $		&	    $						    $		&	    -	       \\
     NGC 4751	  	*	    &		$	 26.3		$	    &		$	   14	    ^{+1}	_{-1}		$	\quad		$	 (14)		$	    &		$	  355	   ^{+18}      _{-18}		$	    &	     R+13	    &	     yes?	    &	   no		    &		$					    $	    \quad	    $			    $		&	    $						    $		&	    -	       \\
     NGC 4826	  		    &		$	  7.3		$	    &		$	0.016	^{+0.004}   _{-0.004}		$	\quad		$			$	    &		$	   91	    ^{+5}	_{-5}		$	    &	     GS13	    &	      no?	    &	   no		    &		$					    $	    \quad	    $			    $		&	    $						    $		&	    -	       \\
     NGC 4889	  	*	    &		$	103.2		$	    &		$	  210	  ^{+160}     _{-160}		$	\quad		$			$	    &		$	  347	   ^{+17}      _{-17}		$	    &	     GS13	    &	      yes	    &	   no		    &		$	597	^{+176}     _{-176}	    $	    \quad	    $			    $		&	    $						    $		&	    -	       \\
     NGC 4945	  		    &		$	  3.8		$	    &		$	0.014	^{+0.014}   _{-0.007}		$	\quad		$			$	    &		$	  100	    ^{+5}	_{-5}		$	    &	     GS13	    &	      no?	    &	  yes		    &		$					    $	    \quad	    $			    $		&	    $						    $		&	    -	       \\
     NGC 5077	  		    &		$	 41.2		$	    &		$	  7.4	  ^{+4.7}	_{-3.0}		$	\quad		$			$	    &		$	  255	   ^{+13}      _{-13}		$	    &	     GS13	    &	      yes	    &	   no		    &		$					    $	    \quad	    $			    $		&	    $						    $		&	    -	       \\
     NGC 5128	  		    &		$	  3.8		$	    &		$	 0.45	 ^{+0.17}     _{-0.10}		$	\quad		$			$	    &		$	  120	    ^{+6}	_{-6}		$	    &	     GS13	    &	      no?	    &	   no		    &		$					    $	    \quad	    $			    $		&	    $						    $		&	    -	       \\
     NGC 5328	  	*	    &		$	 64.1		$	    &		$	   47	   ^{+19}	_{-9}		$	\quad		$      (0.48)		$	    &		$	  333	   ^{+17}      _{-17}		$	    &	     R+13	    &	      yes	    &	   no		    &		$	270	 ^{+41}      _{-41}	    $	    \quad	    $	    (557)	    $		&	    $						    $		&	    -	       \\
     NGC 5516	  	*	    &		$	 58.4		$	    &		$	   33	    ^{+3}	_{-2}		$	\quad		$	 (32)		$	    &		$	  328	   ^{+16}      _{-16}		$	    &	     R+13	    &	      yes	    &	   no		    &		$	135	 ^{+31}      _{-31}	    $	    \quad	    $	    (145)	    $		&	    $						    $		&	    -	       \\
     NGC 5576	  		    &		$	 24.8		$	    &		$	  1.6	  ^{+0.3}     _{-0.4}		$	\quad		$			$	    &		$	  171	    ^{+9}	_{-9}		$	    &	     GS13	    &	       no	    &	   no		    &		$					    $	    \quad	    $			    $		&	    $						    $		&	 SLOW	       \\
     NGC 5813	  		    &		$	 31.3		$	    &		$	  6.8	  ^{+0.7}     _{-0.7}		$	\quad		$			$	    &		$	  239	   ^{+12}      _{-12}		$	    &	     GS13	    &	      yes	    &	   no		    &		$	  6	  ^{+1}       _{-1}	    $	    \quad	    $			    $		&	    $		9	^{+5}	    _{-5}	    $		&	 SLOW	       \\
     NGC 5845	  		    &		$	 25.2		$	    &		$	  2.6	  ^{+0.4}     _{-1.5}		$	\quad		$			$	    &		$	  238	   ^{+12}      _{-12}		$	    &	     GS13	    &	       no	    &	   no		    &		$					    $	    \quad	    $			    $		&	    $						    $		&	 FAST	       \\
     NGC 5846	  		    &		$	 24.2		$	    &		$	   11	    ^{+1}	_{-1}		$	\quad		$			$	    &		$	  237	   ^{+12}      _{-12}		$	    &	     GS13	    &	      yes	    &	   no		    &		$	 23	  ^{+3}       _{-3}	    $	    \quad	    $			    $		&	    $						    $		&	 SLOW	       \\
     NGC 6086	  	*	    &		$	138.0		$	    &		$	   37	   ^{+18}      _{-11}		$	\quad		$			$	    &		$	  318	   ^{+16}      _{-16}		$	    &	     GS13	    &	      yes	    &	   no		    &		$	 76	 ^{+18}      _{-18}	    $	    \quad	    $			    $		&	    $						    $		&	    -	       \\
     NGC 6251	  		    &		$	104.6		$	    &		$	  5.9	    ^{+2.0}	_{-2.0}		$	\quad		$			$	    &		$	  311	   ^{+16}      _{-16}		$	    &	     GS13	    &	     yes?	    &	   no		    &		$					    $	    \quad	    $			    $		&	    $						    $		&	    -	       \\
     NGC 6264	  		    &		$	146.3		$	    &		$	0.305	^{+0.004}   _{-0.004}		$	\quad		$			$	    &		$	  159	    ^{+8}	_{-8}		$	    &	     GS13	    &	      no?	    &	   no		    &		$					    $	    \quad	    $			    $		&	    $						    $		&	    -	       \\
     NGC 6323	  		    &		$	112.4		$	    &		$	  0.100	^{+0.001}   _{-0.001}		$	\quad		$			$	    &		$	  159	    ^{+8}	_{-8}		$	    &	     GS13	    &	      no?	    &	  yes		    &		$					    $	    \quad	    $			    $		&	    $						    $		&	    -	       \\
     NGC 6861	  	*	    &		$	 27.3		$	    &		$	   20	    ^{+2}	_{-2}		$	\quad		$	 (22)		$	    &		$	  389	   ^{+19}      _{-19}		$	    &	     R+13	    &	     yes?	    &	   no		    &		$					    $	    \quad	    $			    $		&	    $						    $		&	    -	       \\
     NGC 7052	  		    &		$	 66.4		$	    &		$	  3.7	  ^{+2.6}     _{-1.5}		$	\quad		$			$	    &		$	  277	   ^{+14}      _{-14}		$	    &	     GS13	    &	      yes	    &	   no		    &		$					    $	    \quad	    $			    $		&	    $						    $		&	    -	       \\
     NGC 7582	  		    &		$	 22.0		$	    &		$	 0.55	 ^{+0.26}    _{-0.19}		$	\quad		$			$	    &		$	  156	    ^{+8}	_{-8}		$	    &	     GS13	    &	       no	    &	  yes		    &		$					    $	    \quad	    $			    $		&	    $						    $		&	    -	       \\
     NGC 7619	  	*	    &		$	 51.5		$	    &		$	   25	    ^{+3}	_{-8}		$	\quad		$	(4.2)		$	    &		$	  292	   ^{+15}      _{-15}		$	    &	     R+13	    &	      yes	    &	   no		    &		$	134	 ^{+28}      _{-28}	    $	    \quad	    $	    (232)	    $		&	    $	       46      ^{+27}	   _{-27}	    $		&	    -	       \\
     NGC 7768	  	*	    &		$	112.8		$	    &		$	   13	    ^{+5}	_{-4}		$	\quad		$			$	    &		$	  257	   ^{+13}      _{-13}		$	    &	     GS13	    &	      yes	    &	   no		    &		$	 25	  ^{+5}       _{-5}	    $	    \quad	    $			    $		&	    $						    $		&	    -	       \\
     UGC 3789	  		    &		$	 48.4		$	    &		$	0.108	^{+0.005}   _{-0.005}		$	\quad		$			$	    &		$	  107	    ^{+5}	_{-5}		$	    &	     GS13	    &	      no?	    &	  yes		    &		$					    $	    \quad	    $			    $		&	    $						    $		&	    -	       \\
\hline
\end{tabular}
\end{center}
\end{table*}

\subsection{Dark matter}
\label{sec:mbh}
The 10 black hole masses from \cite{rusli2013bhmassesDM} 
-- not to be confused with the different 10 galaxies with central mass deficits from \cite{rusli2013}
that are in common with \cite{dullograham2014cores} -- 
were computed by taking into account the effects of dark matter.
For these 10 galaxies, \cite{rusli2013bhmassesDM} also published black hole masses estimated without 
the inclusion of dark matter halos.
Among the 78 black hole masses reported by \cite{grahamscott2013},
only 8 had dark matter included in their derivation, 
and no dark matter halo was included by \cite{greenhill2003} in their black hole mass estimate. \\
The majority\footnote{Stellar mass deficits for IC 1459, 
NGC 3379, NGC 4374 and NGC 4261 were estimated by \cite{rusli2013} with single-component dynamical modelling, i.e. without dark matter.} 
of the 23 stellar mass deficits from \cite{rusli2013} were derived from 
their analysis which incorporated dark matter to obtain the central 
mass-to-light ratios.
However, \cite{rusli2013} did not publish the corresponding stellar mass deficits for the no-dark-matter case. 
Therefore, the sample of 89 galaxies that we use in our analysis contains 18 black hole masses estimated with the inclusion of a dark matter halo
and the 23 stellar mass deficits published by \cite{rusli2013}.\\
We have already shown in Section \ref{sec:data} that the stellar mass deficits measured by \cite{dullograham2014cores}, 
without accounting for dark matter, are in good agreement with the \cite{rusli2013} estimates which accounted for dark matter. 
The slight disagreement observed for the lowest and highest mass deficits 
(see Figure \ref{fig:cfr}) does not significantly affect 
the conclusions of our analysis.
However, one could wonder whether our results change when using exclusively 
black hole masses and stellar mass deficits derived without the inclusion of dark matter.
To address this question, we derived the no-dark-matter 
stellar mass deficits\footnote{The no-dark-matter stellar mass deficits were calculated as
$M_{\rm def,*}^{\rm noDM} = M_{\rm def,*}^{\rm DM} \cdot [(M/L)^{\rm DM}]^{-1} \cdot (M/L)^{\rm noDM}$, 
where $M_{\rm def,*}^{\rm DM}$ are the mass deficits from \cite{rusli2013}, 
which had dark matter incorporated in their derivation, and $(M/L)^{\rm DM}$ and $(M/L)^{\rm noDM}$ 
are the mass-to-light ratios from \cite{rusli2013bhmassesDM} estimated with and without accounting for dark matter respectively.} for 7 of the 10 galaxies 
whose black hole masses were measured by \cite{rusli2013bhmassesDM}.
We repeated the analysis by 
(\emph{i}) employing for these 7 galaxies the no-dark-matter black hole masses (published by \citealt{rusli2013bhmassesDM}) 
and the no-dark-matter stellar mass deficits (derived by us), 
and (\emph{ii}) excluding the remaining black hole masses estimated with the inclusion of dark matter.
We found that none of our conclusions was affected by this change.

%%%%%%%%%%%%%%%%%%%%%%%%%%%%%%%%%%%%%%%%%%%%%%%%%%%%%%%%%%%%%%%%%%%%%%%%%%%%%%%%%%%%%%%%%%%%%%%%%%%%%
%%%%%%%%%%%%%%%%%%%%%%%%%%%%%%%%%%%%%%%%%%%%%%%%%%%%%%%%%%%%%%%%%%%%%%%%%%%%%%%%%%%%%%%%%%%%%%%%%%%%%
%%%%%%%%%%%%%%%%%%%%%%%%%%%%%%%%%%%%%%%%%%%%%%%%%%%%%%%%%%%%%%%%%%%%%%%%%%%%%%%%%%%%%%%%%%%%%%%%%%%%%
%%%%%%%%%%%%%%%%%%%%%%%%%%%%%%%%%%%%%%%%%%%%%%%%%%%%%%%%%%%%%%%%%%%%%%%%%%%%%%%%%%%%%%%%%%%%%%%%%%%%%
%%%%%%%%%%%%%%%%%%%%%%%%%%%%%%%%%%%%%%%%%%%%%%%%%%%%%%%%%%%%%%%%%%%%%%%%%%%%%%%%%%%%%%%%%%%%%%%%%%%%%

\section{Results}
\label{sec:res}
\begin{figure}
\begin{center}
\includegraphics[width=\columnwidth, trim=0 30 100 0]{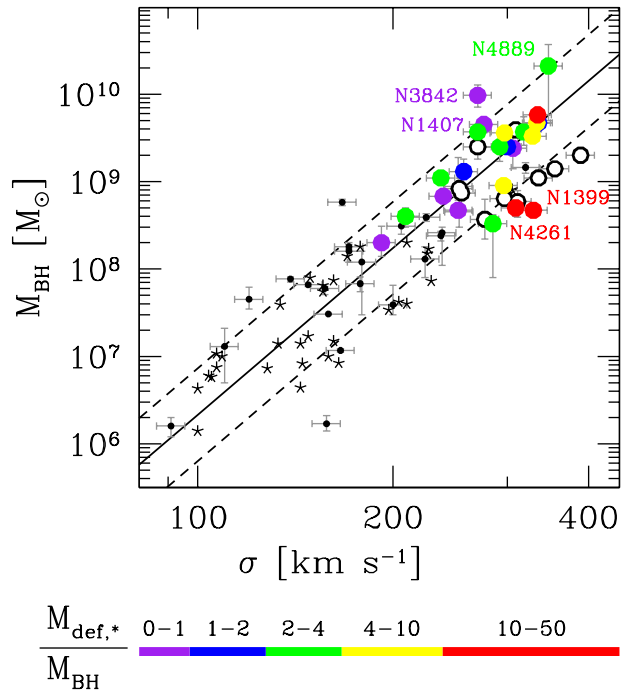}
\end{center}
\caption{$M_{\rm BH} - \sigma$ diagram
for the 89 galaxies presented in Table \ref{tab:data}.
Core-S\'ersic galaxies are colour coded according to their $M_{\rm def,*} / M_{\rm BH}$ ratio.
If no $M_{\rm def,*}$ estimate is available, they appear as open circles.
Unbarred S\'ersic galaxies are represented with (small) black dots 
and barred S\'ersic galaxies with starred symbols.
Errors bars are reported only for unbarred galaxies used to derive Equation \ref{eq:ms}.
The black solid line shows the OLS($\sigma | M_{\rm BH}$) linear regression 
for all non-barred galaxies 
and the black dashed lines mark the associated total rms scatter ($\Delta = 0.53$) in the 
$\log(M_{\rm BH})$ direction.}
\label{fig:mdef}
\end{figure}

\begin{figure}
\begin{center}
\includegraphics[width=\columnwidth, trim=0 30 0 0]{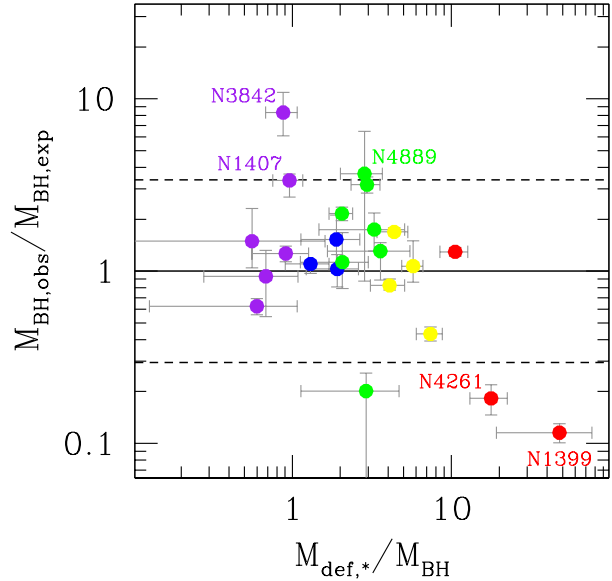}
\end{center}
\caption{Vertical offset from the $M_{\rm BH} - \sigma$ relation versus the $M_{\rm def,*} / M_{\rm BH}$ ratio.
Symbols are colour coded according to Figure \ref{fig:mdef}.
The vertical error bars represent the uncertainty on $M_{\rm BH}$.
The horizontal solid line is equivalent to a zero vertical offset from the \emph{expected} mass 
($M_{\rm BH,obs} / M_{\rm BH,exp}=1$)
and the horizontal dashed lines show the total rms scatter ($\Delta = 0.53$) 
of the OLS($\sigma | M_{\rm BH}$) linear regression 
in the $\log(M_{\rm BH})$ direction.}
\label{fig:offs}
\end{figure}
In Figure \ref{fig:mdef}, we show the updated $M_{\rm BH} - \sigma$ diagram for the 89 galaxies listed in 
Table \ref{tab:data}.
Core-S\'ersic galaxies are colour coded according to their $M_{\rm def,*} / M_{\rm BH}$ ratio
(or, if no $M_{\rm def,*}$ estimate is available, they appear as empty symbols\footnote{The 10 empty symbols
refer to 6 suspected, plus 4 apparent core-S\'ersic galaxies.}).
It is immediately evident that the ``overmassive'' black holes 
are not hosted by galaxies with a high $M_{\rm def,*} / M_{\rm BH}$ value.\\
NGC 4889, NGC 3842 and NGC 1407 are the three objects with the largest positive vertical offset from the 
$M_{\rm BH} - \sigma$ correlation. 
Contrary to expectations, these three galaxies have a small $M_{\rm def,*} / M_{\rm BH}$ ratio, 
consistent with $\sim$1--2 major dry merger events \citep{merritt2006}. \\
Remarkably, NGC 4261 and NGC 1399 -- the central galaxy in the Fornax cluster -- 
{which are two of the three galaxies} with
$M_{\rm def,*} / M_{\rm BH}>10$ (red symbols in Figure \ref{fig:mdef}), 
display a negative vertical offset from the correlation\footnote{Although we have used the black hole mass 
for NGC 1399 from \cite{gebhardt2007}, we note that \cite{houghton2006} had reported a value twice as large 
($\sim$$10^9 \rm~M_{\odot}$).
Nevertheless, this is still too low to yield a positive offset for this galaxy in Figure \ref{fig:mdef}.}. 
While the offset of NGC 1399 and NGC 4261 is at odds with predictions from semi-analytical models 
(see Section \ref{sec:intro}),
their large stellar deficits might be due to the effects of a recoiling black hole 
(see also \citealt{dullograham2014cores,lena2014}).
A recoiling black hole is the final product of a coalesced black hole binary 
after the anisotropic emission of gravitational waves, which imparts a net impulse -- a kick -- to the remnant black hole
\citep{bekenstein1973,fitchettdetweiler1984,favata2004,holleybockelmann2008,batcheldor2010}. 
The kicked black hole oscillates about the centre of the newly merged galaxy with decreasing amplitude, 
transferring kinetic energy to the stars and thus further lowering the core density
\citep{redmountrees1989,merritt2004,boylan-kolchin2004}.
Kick-induced partially depleted cores can be as large as $M_{\rm def,*} \sim (4-5) M_{\rm BH}$
\citep{gualandrismerritt2008} and could complicate the use of central mass deficits as a tracer of dry galaxy mergers.
However, they don't explain the low $M_{\rm def,*} / M_{\rm BH}$ ratios observed in the ``overmassive'' black hole sample. \\
In Figure \ref{fig:offs}, we plot the vertical offset
from the $M_{\rm BH} - \sigma$ relation versus the $M_{\rm def,*} / M_{\rm BH}$ ratio.
The vertical offset is defined as $\log( M_{\rm BH,obs} / M_{\rm BH,exp} )$, 
where $M_{\rm BH,obs}$ is the \emph{observed} black hole mass 
and $M_{\rm BH,exp}$ is the black hole mass \emph{expected} from the galaxy velocity dispersion
using an OLS($\sigma | M_{\rm BH}$) linear regression\footnote{See \citeauthor{grahamscott2013} 
(\citeyear{grahamscott2013}, their Section 3.1)
for a discussion on the choice of an ordinary least-squares (OLS) regression of the abscissa on the ordinate.
Their OLS($\sigma | M_{\rm BH}$) linear regression for unbarred galaxies 
($\log(M_{\rm BH,exp}/{\rm M_\odot}) = (8.22 \pm 0.05) + (5.53 \pm 0.34) \times \log(\sigma/200 \rm~km~s^{-1})$)
is consistent within the over-lapping 1$\sigma$ uncertainties.
It is however beyond the scope of this paper to repeat the same detailed analysis presented by \cite{grahamscott2013}.} 
for all non-barred\footnote{As noted in Section \ref{sec:intro}, barred galaxies tend to be offset from non-barred galaxies 
in the $M_{\rm BH} - \sigma$ diagram.} galaxies: 
\begin{multline}
\log \biggl(\frac{M_{\rm BH,exp}}{{\rm M_\odot}}\biggr) = \\ = (8.24 \pm 0.10) + (6.34 \pm 0.80) \times \log \biggl(\frac{\sigma}{200 \rm~km~s^{-1}}\biggr) .
\label{eq:ms}\end{multline}
Clearly, there is no positive trend in Figure \ref{fig:offs}.
The significance of a correlation is rejected by a Spearman's test 
{(Spearman's correlation coefficient $r_{\rm s} = -0.33$, likelihood of the correlation occuring by chance $P > 5\%$).}
We conclude that no positive correlation is observed between the vertical offset from the $M_{\rm BH} - \sigma$ relation 
and the $M_{\rm def,*} / M_{\rm BH}$ ratio. \\
Repeating the analysis using only the \cite{rusli2013} mass deficits, 
i.e. without computing 10 weighted arithmetic means for the galaxies in common with \cite{dullograham2014cores},
gives the same conclusion.
Similarly, the same conclusion is reached when using only the 10 \cite{dullograham2014cores} derived mass deficits. \\

\begin{figure}
\begin{center}
\includegraphics[width=\columnwidth, trim=0 30 0 0]{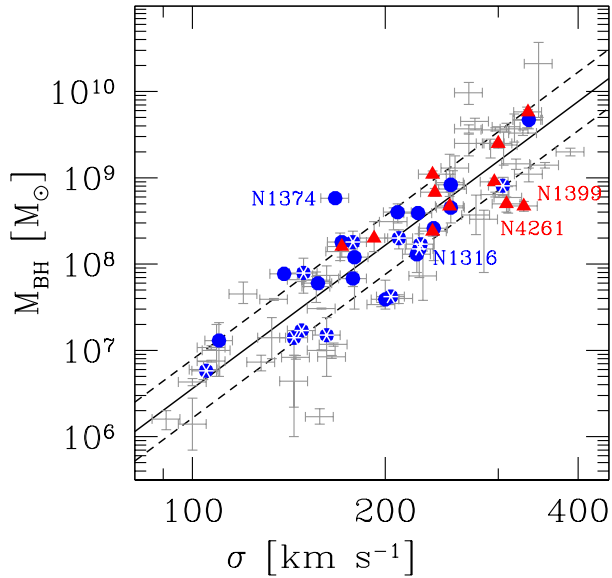}
\end{center}
\caption{Fast (blue circles) and slow (red triangles) rotators 
in the $M_{\rm BH} - \sigma$ diagram.
Starred symbols mark barred galaxies.
The black solid and dashed lines are the same as in Figure \ref{fig:mdef}.}
\label{fig:kinem}
\end{figure}

In Figure \ref{fig:kinem} we show the distribution of fast and slow rotators in the $M_{\rm BH} - \sigma$ diagram.
Our aim is to check whether the two populations are vertically offset from each other,
in the sense that wet mergers can create fast rotating discs,
while dry mergers can increase the black hole mass but not the velocity dispersion.
Since the work of \cite{graham2008FP,graham2008bar}, see also \cite{hu2008},
we know that barred galaxies tend to be offset rightward 
from the $M_{\rm BH} - \sigma$ correlation defined by non-barred galaxies. 
It is therefore crucial to exclude the barred galaxies from the following analysis,
to avoid biasing the results. 
We follow \cite{grahamscott2013} in using the BCES code from \cite{akritasbershady1996}
to obtain four different linear regressions for both the (unbarred) fast and slow rotators.
The results are shown in the first part of Table \ref{tab:lreg}.
Regardless of the linear regression method used, 
the best-fit slopes and intercepts of fast and slow rotators are consistent with each other within the $1\sigma$ uncertainty.
To test the robustness of our results, 
we repeated the linear regression analysis excluding the most deviating data points:
one fast rotator with a positive vertical offset (NGC 1374)
and two slow rotators with a negative vertical offset (NGC 1399 and NGC 4261).
The second part of Table \ref{tab:lreg} reports the new values of the best-fit slopes and intercepts, 
which remain consistent with each other.\\

\begin{table}
\centering
\caption{Linear regression analysis for the populations of unbarred fast and slow rotators.}
\begin{tabular}{lcccc}
\hline
\hline
  & \multicolumn{2}{|c|}{Slow rot.}  & \multicolumn{2}{|c|}{Fast rot.} \\
\hline
 & \multicolumn{4}{|c|}{$\log[M_{\rm BH}/{\rm M_\odot}] = \alpha + \beta\log[{\sigma/(200{\rm~km~s^{-1}})}]$} \\ [0.3em]
{\bf Regression} & $\boldsymbol \beta$ & $\boldsymbol \alpha$ & $\boldsymbol \beta$ & $\boldsymbol \alpha$ \\ 
\hline 
OLS($M_{\rm BH} | \sigma$) & $3.7 \pm 1.1$ &  $8.40 \pm 0.08$   &  $4.4 \pm 0.6$ & $8.33 \pm 0.09$ \\
OLS($\sigma | M_{\rm BH}$) & $6.8 \pm 1.7$ &  $8.1  \pm 0.3 $   &  $5.9 \pm 1.0$ & $8.3  \pm 0.1 $ \\
Bisector                   & $4.8 \pm 1.0$ &  $8.3  \pm 0.1 $   &  $5.1 \pm 0.5$ & $8.33 \pm 0.09$ \\
Orthogonal                 & $6.7 \pm 1.6$ &  $8.1  \pm 0.3 $   &  $5.8 \pm 1.0$ & $8.3  \pm 0.1 $ \\
\hline
\multicolumn{5}{|c|}{\emph{Excluding NGC 1374, NGC 1399 and NGC 4261.}} \\ [0.3em]
OLS($M_{\rm BH} | \sigma$) & $5.3 \pm 0.8$ &  $8.36 \pm 0.09$   &  $4.7 \pm 0.6$ & $8.27 \pm 0.08$ \\
OLS($\sigma | M_{\rm BH}$) & $6.2 \pm 0.9$ &  $8.3  \pm 0.1 $   &  $5.4 \pm 0.8$ & $8.28 \pm 0.08$ \\
Bisector                   & $5.7 \pm 0.8$ &  $8.3  \pm 0.1 $   &  $5.0 \pm 0.6$ & $8.27 \pm 0.08$ \\
Orthogonal                 & $6.1 \pm 0.9$ &  $8.3  \pm 0.1 $   &  $5.4 \pm 0.8$ & $8.28 \pm 0.08$ \\
\hline
\end{tabular}
\label{tab:lreg} 
\end{table}

%%%%%%%%%%%%%%%%%%%%%%%%%%%%%%%%%%%%%%%%%%%%%%%%%%%%%%%%%%%%%%%%%%%%%%%%%%%%%%%%%%%%%%%%%%%%%%%%%%%%%
%%%%%%%%%%%%%%%%%%%%%%%%%%%%%%%%%%%%%%%%%%%%%%%%%%%%%%%%%%%%%%%%%%%%%%%%%%%%%%%%%%%%%%%%%%%%%%%%%%%%%
%%%%%%%%%%%%%%%%%%%%%%%%%%%%%%%%%%%%%%%%%%%%%%%%%%%%%%%%%%%%%%%%%%%%%%%%%%%%%%%%%%%%%%%%%%%%%%%%%%%%%
%%%%%%%%%%%%%%%%%%%%%%%%%%%%%%%%%%%%%%%%%%%%%%%%%%%%%%%%%%%%%%%%%%%%%%%%%%%%%%%%%%%%%%%%%%%%%%%%%%%%%
%%%%%%%%%%%%%%%%%%%%%%%%%%%%%%%%%%%%%%%%%%%%%%%%%%%%%%%%%%%%%%%%%%%%%%%%%%%%%%%%%%%%%%%%%%%%%%%%%%%%%

\section{Discussion and conclusions}
\label{sec:concl}
The presence of a central, supermassive black hole, coupled with the scarcity of binary supermassive black hole systems,
suggests that the progenitor black holes have coalesced in most merged galaxies.
They can do this by transferring their orbital angular momentum to the stars near the centre of their host galaxy 
and thereby evacuating the core.
If a galaxy's $M_{\rm def,*} / M_{\rm BH}$ ratio is a proxy for its equivalent number of major dry merger events
since its last wet merger (e.g. \citealt{merritt2006}),
then our analysis (see Figures \ref{fig:mdef} and \ref{fig:offs}) reveals that the apparent ``overmassive'' outliers 
at the high-mass end of the $M_{\rm BH} - \sigma$ diagram 
are galaxies that have undergone the lowest degree of such recent dry merging.
Although a final major wet merger may contribute to their low $M_{\rm def,*} / M_{\rm BH}$ ratio,
these galaxies are among the most massive early-type galaxies in the local Universe
and they reside in the central regions of galaxy clusters, 
where wet major mergers are unlikely to occur (e.g. \citealt{frasermckelvie2014}) 
due to prior ram pressure stripping of gas from infalling galaxies
\citep{boselligavazzi2006,haines2013,boselli2014HRS,boselli2014GUVICS}.
That is, the ``overmassive'' black holes in central cluster galaxies 
cannot be explained by a large number of dissipationless mergers growing 
the black hole mass at a fixed galaxy velocity dispersion. \\
In addition to this,  
no significant offset is observed between the (unbarred) populations of 
fast and slow rotators in the $M_{\rm BH} - \sigma$ diagram (see Table \ref{tab:lreg}),
contrary to what is expected if fast and slow rotators are, in general, the products of wet and dry mergers respectively. 
This is because dry mergers will increase the black hole mass, but are said not to increase the velocity dispersion.
This result is also in broad agreement with the observation that the (unbarred) S\'ersic and core-S\'ersic galaxies
follow the same $M_{\rm BH} - \sigma$ relation \citep{grahamscott2013}. 
Our results appear consistent with studies of luminous elliptical galaxies which have shown that 
the galaxy luminosity scales 
with the velocity dispersion
\citep{schechter1980,malumuthkirshner1981,vonderlinden2007,lauer2007lumell,bernardi2007,liu2008},
i.e. the velocity dispersion appears not to completely saturate but rather still increases
with increasing galaxy luminosity,
contrary to what one would predict if these galaxies were built only by dry mergers on parabolic orbits. \\
An alternative possibility for the central cluster galaxies may be that they experience minor dry merger events
that do not bring in a massive black hole but rather stars, and nuclear star clusters,
which may partly or fully refill a depleted galaxy core.
However, simulations are needed to verify whether, in a $\Lambda$CDM cosmology, 
the extent of minor dry mergers experienced by a central cluster galaxy in late cosmic times 
can supply enough stellar mass ($\sim 10^9 - 10^{10} \rm~M_\odot$) to replenish the galaxy's core.\\
Eventually, one should also consider the possibility that some of the overmassive black holes might have had their masses overestimated.
Past studies have demonstrated the importance of resolving the black hole 
sphere-of-influence\footnote{The sphere-of-influence is the region of space within which the gravitational potential 
of the black hole dominates over that of the surrounding stars.} 
when measuring a black hole mass, to avoid systematic errors or even spurious detections
(e.g. \citealt{ferraresemerritt2000,merrittferrarese2001sphinfl,merrittferrarese2001ms,valluri2004,ferrareseford2005}). 
\cite{merritt2013book} cautions against the use of black hole mass measurements obtained from stellar-dynamical data sets.
His Figure 2.5 points out that no more than three galaxies -- all belonging to the Local Group -- 
have been observed with enough spatial resolution 
to exhibit a \emph{prima facie} convincing Keplerian rise in their central stellar velocities.  
At the same time, gas kinematics can have motions not solely due to the gravitational potential of the black hole. 
For example, 
\cite{mazzalay2014} showed that the gas dynamics in the innermost parsecs of spiral galaxies
is typically far from simple circular motion.  
One possible example of such an overestimated black hole may be that reported by \cite{vandenbosch2012} 
for the galaxy NGC 1277 ($M_{\rm BH} = 1.7 \times 10^{10} \rm~M_\odot$).
In fact, upon re-analyzing the same data, \cite{emsellem2013} showed that a model with a 2 times smaller black hole mass
provides an equally good fit to the observed kinematics, and emphasized the need for higher spatial resolution spectroscopic data.

%%%%%%%%%%%%%%%%%%%%%%%%%%%%%%%%%%%%%%%%%%%%%%%%%%%%%%%%%%%%%%%%%%%%%%%%%%%%%%%%%%%%%%%%%%%%%%%%%%%%%
%%%%%%%%%%%%%%%%%%%%%%%%%%%%%%%%%%%%%%%%%%%%%%%%%%%%%%%%%%%%%%%%%%%%%%%%%%%%%%%%%%%%%%%%%%%%%%%%%%%%%
%%%%%%%%%%%%%%%%%%%%%%%%%%%%%%%%%%%%%%%%%%%%%%%%%%%%%%%%%%%%%%%%%%%%%%%%%%%%%%%%%%%%%%%%%%%%%%%%%%%%%
%%%%%%%%%%%%%%%%%%%%%%%%%%%%%%%%%%%%%%%%%%%%%%%%%%%%%%%%%%%%%%%%%%%%%%%%%%%%%%%%%%%%%%%%%%%%%%%%%%%%%
%%%%%%%%%%%%%%%%%%%%%%%%%%%%%%%%%%%%%%%%%%%%%%%%%%%%%%%%%%%%%%%%%%%%%%%%%%%%%%%%%%%%%%%%%%%%%%%%%%%%%

\section*{Acknowledgments}
GS would like to acknowledge the valuable feedback provided by David Merritt and Luca Ciotti on an early version of the manuscript.
GS thanks Gonzalo D\'iaz and Bililign Dullo for useful discussions. % karl, nicola? 
This research was supported by Australian Research Council funding through grants DP110103509 and FT110100263.
This research has made use of the GOLDMine website \citep{goldmine} and 
the NASA/IPAC Extragalactic Database (NED) which is operated 
by the Jet Propulsion Laboratory, California Institute of Technology, 
under contract with the National Aeronautics and Space Administration.
We wish to thank an anonymous referee whose criticism helped improving the manuscript.

\bibliography{SMBHbibliography}

\begin{thebibliography}{100}
\expandafter\ifx\csname natexlab\endcsname\relax\def\natexlab#1{#1}\fi

\bibitem[{{Akritas} \& {Bershady}(1996)}]{akritasbershady1996}
{Akritas} M.~G., {Bershady} M.~A., 1996, \apj, 470, 706

\bibitem[{{Athanassoula} \& {Beaton}(2006)}]{athanassoulabeaton2006m31}
{Athanassoula} E., {Beaton} R.~L., 2006, \mnras, 370, 1499

\bibitem[{{Batcheldor} {et~al}\mbox{.}(2010){Batcheldor}, {Robinson}, {Axon},
  {Perlman}, \& {Merritt}}]{batcheldor2010}
{Batcheldor} D., {Robinson} A., {Axon} D.~J., {Perlman} E.~S., {Merritt} D.,
  2010, \apjl, 717, L6

\bibitem[{{Beaton} {et~al}\mbox{.}(2007){Beaton}, {Majewski}, {Guhathakurta},
  {Skrutskie}, {Cutri}, {Good}, {Patterson}, {Athanassoula}, \&
  {Bureau}}]{beaton2007m31}
{Beaton} R.~L. {et~al.}, 2007, \apjl, 658, L91

\bibitem[{{Begelman}, {Blandford} \& {Rees}(1980){Begelman}, {Blandford}, \&
  {Rees}}]{begelman1980}
{Begelman} M.~C., {Blandford} R.~D., {Rees} M.~J., 1980, \nat, 287, 307

\bibitem[{{Bekenstein}(1973)}]{bekenstein1973}
{Bekenstein} J.~D., 1973, \apj, 183, 657

\bibitem[{{Bernardi} {et~al}\mbox{.}(2007){Bernardi}, {Hyde}, {Sheth},
  {Miller}, \& {Nichol}}]{bernardi2007}
{Bernardi} M., {Hyde} J.~B., {Sheth} R.~K., {Miller} C.~J., {Nichol} R.~C.,
  2007, \aj, 133, 1741

\bibitem[{{Boselli} {et~al}\mbox{.}(2014{\natexlab{a}}){Boselli}, {Cortese},
  {Boquien}, {Boissier}, {Catinella}, {Gavazzi}, {Lagos}, \&
  {Saintonge}}]{boselli2014HRS}
{Boselli} A., {Cortese} L., {Boquien} M., {Boissier} S., {Catinella} B.,
  {Gavazzi} G., {Lagos} C., {Saintonge} A., 2014{\natexlab{a}}, \aap, 564, A67

\bibitem[{{Boselli} \& {Gavazzi}(2006)}]{boselligavazzi2006}
{Boselli} A., {Gavazzi} G., 2006, \pasp, 118, 517

\bibitem[{{Boselli} {et~al}\mbox{.}(2014{\natexlab{b}}){Boselli}, {Voyer},
  {Boissier}, {Cucciati}, {Consolandi}, {Cortese}, {Fumagalli}, {Gavazzi},
  {Heinis}, {Roehlly}, \& {Toloba}}]{boselli2014GUVICS}
{Boselli} A. {et~al.}, 2014{\natexlab{b}}, ArXiv e-prints

\bibitem[{{Boylan-Kolchin}, {Ma} \& {Quataert}(2004){Boylan-Kolchin}, {Ma}, \&
  {Quataert}}]{boylan-kolchin2004}
{Boylan-Kolchin} M., {Ma} C.-P., {Quataert} E., 2004, \apjl, 613, L37

\bibitem[{{Brown} {et~al}\mbox{.}(2013){Brown}, {Valluri}, {Shen}, \&
  {Debattista}}]{brown2013}
{Brown} J.~S., {Valluri} M., {Shen} J., {Debattista} V.~P., 2013, \apj, 778,
  151

\bibitem[{{Burke-Spolaor}(2011)}]{burke2011}
{Burke-Spolaor} S., 2011, \mnras, 410, 2113

\bibitem[{{Ciotti}, {Lanzoni} \& {Volonteri}(2007){Ciotti}, {Lanzoni}, \&
  {Volonteri}}]{ciotti2007}
{Ciotti} L., {Lanzoni} B., {Volonteri} M., 2007, \apj, 658, 65

\bibitem[{{Ciotti}, {Ostriker} \& {Proga}(2010){Ciotti}, {Ostriker}, \&
  {Proga}}]{ciotti2010}
{Ciotti} L., {Ostriker} J.~P., {Proga} D., 2010, \apj, 717, 708

\bibitem[{{Ciotti} \& {van Albada}(2001)}]{ciottivanalbada2001}
{Ciotti} L., {van Albada} T.~S., 2001, \apjl, 552, L13

\bibitem[{{Colpi}(2014)}]{colpi2014}
{Colpi} M., 2014, \ssr

\bibitem[{{de Souza}, {Gadotti} \& {dos Anjos}(2004){de Souza}, {Gadotti}, \&
  {dos Anjos}}]{desouza2004}
{de Souza} R.~E., {Gadotti} D.~A., {dos Anjos} S., 2004, \apjs, 153, 411

\bibitem[{{D'Onofrio}(2001)}]{donofrio2001}
{D'Onofrio} M., 2001, \mnras, 326, 1517

\bibitem[{{Dotti} {et~al}\mbox{.}(2009){Dotti}, {Montuori}, {Decarli},
  {Volonteri}, {Colpi}, \& {Haardt}}]{dotti2009}
{Dotti} M., {Montuori} C., {Decarli} R., {Volonteri} M., {Colpi} M., {Haardt}
  F., 2009, \mnras, 398, L73

\bibitem[{{Dotti}, {Sesana} \& {Decarli}(2012){Dotti}, {Sesana}, \&
  {Decarli}}]{dotti2012}
{Dotti} M., {Sesana} A., {Decarli} R., 2012, Advances in Astronomy, 2012

\bibitem[{{Dullo} \& {Graham}(2014)}]{dullograham2014cores}
{Dullo} B.~T., {Graham} A.~W., 2014, \mnras, 444, 2700

\bibitem[{{Elmegreen}, {Chromey} \& {Johnson}(1995){Elmegreen}, {Chromey}, \&
  {Johnson}}]{elmegreen1995m81}
{Elmegreen} D.~M., {Chromey} F.~R., {Johnson} C.~O., 1995, \aj, 110, 2102

\bibitem[{{Emsellem}(2013)}]{emsellem2013}
{Emsellem} E., 2013, \mnras, 433, 1862

\bibitem[{{Emsellem} {et~al}\mbox{.}(2011){Emsellem}, {Cappellari},
  {Krajnovi{\'c}}, {Alatalo}, {Blitz}, {Bois}, {Bournaud}, {Bureau}, {Davies},
  {Davis}, {de Zeeuw}, {Khochfar}, {Kuntschner}, {Lablanche}, {McDermid},
  {Morganti}, {Naab}, {Oosterloo}, {Sarzi}, {Scott}, {Serra}, {van de Ven},
  {Weijmans}, \& {Young}}]{atlas3dIII}
{Emsellem} E. {et~al.}, 2011, \mnras, 414, 888

\bibitem[{{Emsellem} {et~al}\mbox{.}(2008){Emsellem}, {Cappellari},
  {Krajnovi{\'c}}, {van de Ven}, {Bacon}, {Bureau}, {Davies}, {de Zeeuw},
  {Falc{\'o}n-Barroso}, {Kuntschner}, {McDermid}, {Peletier}, {Sarzi}, \& {van
  den Bosch}}]{emsellem2008fastslow}
{Emsellem} E. {et~al.}, 2008, in IAU Symposium, Vol. 245, IAU Symposium,
  {Bureau} M., {Athanassoula} E., {Barbuy} B., eds., pp. 11--14

\bibitem[{{Erwin} \& {Debattista}(2013)}]{erwindebattista2013}
{Erwin} P., {Debattista} V.~P., 2013, \mnras, 431, 3060

\bibitem[{{Fabbiano} {et~al}\mbox{.}(2011){Fabbiano}, {Wang}, {Elvis}, \&
  {Risaliti}}]{fabbiano2011}
{Fabbiano} G., {Wang} J., {Elvis} M., {Risaliti} G., 2011, \nat, 477, 431

\bibitem[{{Favata}, {Hughes} \& {Holz}(2004){Favata}, {Hughes}, \&
  {Holz}}]{favata2004}
{Favata} M., {Hughes} S.~A., {Holz} D.~E., 2004, \apjl, 607, L5

\bibitem[{{Ferrarese} {et~al}\mbox{.}(2006){Ferrarese}, {C{\^o}t{\'e}},
  {Jord{\'a}n}, {Peng}, {Blakeslee}, {Piatek}, {Mei}, {Merritt},
  {Milosavljevi{\'c}}, {Tonry}, \& {West}}]{ferrarese2006acsvcs}
{Ferrarese} L. {et~al.}, 2006, \apjs, 164, 334

\bibitem[{{Ferrarese} \& {Ford}(2005)}]{ferrareseford2005}
{Ferrarese} L., {Ford} H., 2005, \ssr, 116, 523

\bibitem[{{Ferrarese} \& {Merritt}(2000)}]{ferraresemerritt2000}
{Ferrarese} L., {Merritt} D., 2000, \apjl, 539, L9

\bibitem[{{Fitchett} \& {Detweiler}(1984)}]{fitchettdetweiler1984}
{Fitchett} M.~J., {Detweiler} S., 1984, \mnras, 211, 933

\bibitem[{{Fraser-McKelvie}, {Brown} \& {Pimbblet}(2014){Fraser-McKelvie},
  {Brown}, \& {Pimbblet}}]{frasermckelvie2014}
{Fraser-McKelvie} A., {Brown} M.~J.~I., {Pimbblet} K.~A., 2014, ArXiv e-prints

\bibitem[{{Gavazzi} {et~al}\mbox{.}(2003){Gavazzi}, {Boselli}, {Donati},
  {Franzetti}, \& {Scodeggio}}]{goldmine}
{Gavazzi} G., {Boselli} A., {Donati} A., {Franzetti} P., {Scodeggio} M., 2003,
  \aap, 400, 451

\bibitem[{{Gebhardt} {et~al}\mbox{.}(2000){Gebhardt}, {Bender}, {Bower},
  {Dressler}, {Faber}, {Filippenko}, {Green}, {Grillmair}, {Ho}, {Kormendy},
  {Lauer}, {Magorrian}, {Pinkney}, {Richstone}, \& {Tremaine}}]{gebhardt2000}
{Gebhardt} K. {et~al.}, 2000, \apjl, 539, L13

\bibitem[{{Gebhardt} {et~al}\mbox{.}(2007){Gebhardt}, {Lauer}, {Pinkney},
  {Bender}, {Richstone}, {Aller}, {Bower}, {Dressler}, {Faber}, {Filippenko},
  {Green}, {Ho}, {Kormendy}, {Siopis}, \& {Tremaine}}]{gebhardt2007}
{Gebhardt} K. {et~al.}, 2007, \apj, 671, 1321

\bibitem[{{Graham}(2007)}]{graham2007baas}
{Graham} A., 2007, in Bulletin of the American Astronomical Society, Vol.~39,
  American Astronomical Society Meeting Abstracts, p. 759

\bibitem[{{Graham}(2004)}]{graham2004}
{Graham} A.~W., 2004, \apjl, 613, L33

\bibitem[{{Graham}(2008{\natexlab{a}})}]{graham2008FP}
{Graham} A.~W., 2008{\natexlab{a}}, \apj, 680, 143

\bibitem[{{Graham}(2008{\natexlab{b}})}]{graham2008bar}
{Graham} A.~W., 2008{\natexlab{b}}, \pasa, 25, 167

\bibitem[{{Graham}(2012)}]{graham2012bent}
{Graham} A.~W., 2012, \apj, 746, 113

\bibitem[{{Graham} {et~al}\mbox{.}(2003){Graham}, {Erwin}, {Trujillo}, \&
  {Asensio Ramos}}]{graham2003coresersicmodel}
{Graham} A.~W., {Erwin} P., {Trujillo} I., {Asensio Ramos} A., 2003, \aj, 125,
  2951

\bibitem[{{Graham} \& {Li}(2009)}]{grahamli2009}
{Graham} A.~W., {Li} I.-h., 2009, \apj, 698, 812

\bibitem[{{Graham} {et~al}\mbox{.}(2011){Graham}, {Onken}, {Athanassoula}, \&
  {Combes}}]{graham2011}
{Graham} A.~W., {Onken} C.~A., {Athanassoula} E., {Combes} F., 2011, \mnras,
  412, 2211

\bibitem[{{Graham} \& {Scott}(2013)}]{grahamscott2013}
{Graham} A.~W., {Scott} N., 2013, \apj, 764, 151

\bibitem[{{Graham} \& {Spitler}(2009)}]{grahamspitler2009}
{Graham} A.~W., {Spitler} L.~R., 2009, \mnras, 397, 2148

\bibitem[{{Greenhill} {et~al}\mbox{.}(2003){Greenhill}, {Booth}, {Ellingsen},
  {Herrnstein}, {Jauncey}, {McCulloch}, {Moran}, {Norris}, {Reynolds}, \&
  {Tzioumis}}]{greenhill2003}
{Greenhill} L.~J. {et~al.}, 2003, \apj, 590, 162

\bibitem[{{Gualandris} \& {Merritt}(2008)}]{gualandrismerritt2008}
{Gualandris} A., {Merritt} D., 2008, \apj, 678, 780

\bibitem[{{Guti{\'e}rrez} {et~al}\mbox{.}(2011){Guti{\'e}rrez}, {Erwin},
  {Aladro}, \& {Beckman}}]{gutierrez2011}
{Guti{\'e}rrez} L., {Erwin} P., {Aladro} R., {Beckman} J.~E., 2011, \aj, 142,
  145

\bibitem[{{Haines} {et~al}\mbox{.}(2013){Haines}, {Pereira}, {Smith}, {Egami},
  {Sanderson}, {Babul}, {Finoguenov}, {Merluzzi}, {Busarello}, {Rawle}, \&
  {Okabe}}]{haines2013}
{Haines} C.~P. {et~al.}, 2013, \apj, 775, 126

\bibitem[{{Hartmann} {et~al}\mbox{.}(2013){Hartmann}, {Debattista}, {Cole},
  {Valluri}, {Widrow}, \& {Shen}}]{hartmann2013}
{Hartmann} M., {Debattista} V.~P., {Cole} D.~R., {Valluri} M., {Widrow} L.~M.,
  {Shen} J., 2013, ArXiv e-prints

\bibitem[{{Hirschmann} {et~al}\mbox{.}(2010){Hirschmann}, {Khochfar},
  {Burkert}, {Naab}, {Genel}, \& {Somerville}}]{hirschmann2010}
{Hirschmann} M., {Khochfar} S., {Burkert} A., {Naab} T., {Genel} S.,
  {Somerville} R.~S., 2010, \mnras, 407, 1016

\bibitem[{{Holley-Bockelmann} {et~al}\mbox{.}(2008){Holley-Bockelmann},
  {G{\"u}ltekin}, {Shoemaker}, \& {Yunes}}]{holleybockelmann2008}
{Holley-Bockelmann} K., {G{\"u}ltekin} K., {Shoemaker} D., {Yunes} N., 2008,
  \apj, 686, 829

\bibitem[{{Houghton} {et~al}\mbox{.}(2006){Houghton}, {Magorrian}, {Sarzi},
  {Thatte}, {Davies}, \& {Krajnovi{\'c}}}]{houghton2006}
{Houghton} R.~C.~W., {Magorrian} J., {Sarzi} M., {Thatte} N., {Davies} R.~L.,
  {Krajnovi{\'c}} D., 2006, \mnras, 367, 2

\bibitem[{{Hu}(2008)}]{hu2008}
{Hu} J., 2008, \mnras, 386, 2242

\bibitem[{{Hyde} {et~al}\mbox{.}(2008){Hyde}, {Bernardi}, {Sheth}, \&
  {Nichol}}]{hyde2008}
{Hyde} J.~B., {Bernardi} M., {Sheth} R.~K., {Nichol} R.~C., 2008, \mnras, 391,
  1559

\bibitem[{{Jahnke} \& {Macci{\`o}}(2011)}]{jahnkemaccio2011}
{Jahnke} K., {Macci{\`o}} A.~V., 2011, \apj, 734, 92

\bibitem[{{Jeong} {et~al}\mbox{.}(2007){Jeong}, {Bureau}, {Yi},
  {Krajnovi{\'c}}, \& {Davies}}]{jeong2007}
{Jeong} H., {Bureau} M., {Yi} S.~K., {Krajnovi{\'c}} D., {Davies} R.~L., 2007,
  \mnras, 376, 1021

\bibitem[{{Ju} {et~al}\mbox{.}(2013){Ju}, {Greene}, {Rafikov}, {Bickerton}, \&
  {Badenes}}]{ju2013}
{Ju} W., {Greene} J.~E., {Rafikov} R.~R., {Bickerton} S.~J., {Badenes} C.,
  2013, \apj, 777, 44

\bibitem[{{Komossa} {et~al}\mbox{.}(2003){Komossa}, {Burwitz}, {Hasinger},
  {Predehl}, {Kaastra}, \& {Ikebe}}]{komossa2003}
{Komossa} S., {Burwitz} V., {Hasinger} G., {Predehl} P., {Kaastra} J.~S.,
  {Ikebe} Y., 2003, \apjl, 582, L15

\bibitem[{{Kormendy}, {Bender} \& {Cornell}(2011){Kormendy}, {Bender}, \&
  {Cornell}}]{kormendybender2011}
{Kormendy} J., {Bender} R., {Cornell} M.~E., 2011, \nat, 469, 374

\bibitem[{{Lauer} {et~al}\mbox{.}(2007){Lauer}, {Faber}, {Richstone},
  {Gebhardt}, {Tremaine}, {Postman}, {Dressler}, {Aller}, {Filippenko},
  {Green}, {Ho}, {Kormendy}, {Magorrian}, \& {Pinkney}}]{lauer2007lumell}
{Lauer} T.~R. {et~al.}, 2007, \apj, 662, 808

\bibitem[{{Laurikainen} {et~al}\mbox{.}(2010){Laurikainen}, {Salo}, {Buta},
  {Knapen}, \& {Comer{\'o}n}}]{laurikainen2010}
{Laurikainen} E., {Salo} H., {Buta} R., {Knapen} J.~H., {Comer{\'o}n} S., 2010,
  \mnras, 405, 1089

\bibitem[{{Lena} {et~al}\mbox{.}(2014){Lena}, {Robinson}, {Marconi}, {Axon},
  {Capetti}, {Merritt}, \& {Batcheldor}}]{lena2014}
{Lena} D., {Robinson} A., {Marconi} A., {Axon} D.~J., {Capetti} A., {Merritt}
  D., {Batcheldor} D., 2014, ArXiv e-prints

\bibitem[{{Liu} {et~al}\mbox{.}(2008){Liu}, {Xia}, {Mao}, {Wu}, \&
  {Deng}}]{liu2008}
{Liu} F.~S., {Xia} X.~Y., {Mao} S., {Wu} H., {Deng} Z.~G., 2008, \mnras, 385,
  23

\bibitem[{Liu {et~al}\mbox{.}(2014)Liu, Shen, Bian, Loeb, \&
  Tremaine}]{liu2014}
Liu X., Shen Y., Bian F., Loeb A., Tremaine S., 2014, \apj, 789, 140

\bibitem[{{Malumuth} \& {Kirshner}(1981)}]{malumuthkirshner1981}
{Malumuth} E.~M., {Kirshner} R.~P., 1981, \apj, 251, 508

\bibitem[{{Maness} {et~al}\mbox{.}(2004){Maness}, {Taylor}, {Zavala}, {Peck},
  \& {Pollack}}]{maness2004}
{Maness} H.~L., {Taylor} G.~B., {Zavala} R.~T., {Peck} A.~B., {Pollack} L.~K.,
  2004, \apj, 602, 123

\bibitem[{{Mazzalay} {et~al}\mbox{.}(2014){Mazzalay}, {Maciejewski}, {Erwin},
  {Saglia}, {Bender}, {Fabricius}, {Nowak}, {Rusli}, \&
  {Thomas}}]{mazzalay2014}
{Mazzalay} X. {et~al.}, 2014, \mnras, 438, 2036

\bibitem[{{McConnell} \& {Ma}(2013)}]{mcconnellma2013}
{McConnell} N.~J., {Ma} C.-P., 2013, \apj, 764, 184

\bibitem[{{McNeil-Moylan} {et~al}\mbox{.}(2012){McNeil-Moylan}, {Freeman},
  {Arnaboldi}, \& {Gerhard}}]{mcneilmoylan2012}
{McNeil-Moylan} E.~K., {Freeman} K.~C., {Arnaboldi} M., {Gerhard} O.~E., 2012,
  \aap, 539, A11

\bibitem[{{Merritt}(2006{\natexlab{a}})}]{merritt2006RPP}
{Merritt} D., 2006{\natexlab{a}}, Reports on Progress in Physics, 69, 2513

\bibitem[{{Merritt}(2006{\natexlab{b}})}]{merritt2006}
{Merritt} D., 2006{\natexlab{b}}, \apj, 648, 976

\bibitem[{{Merritt}(2013{\natexlab{a}})}]{merritt2013book}
{Merritt} D., 2013{\natexlab{a}}, {Dynamics and Evolution of Galactic Nuclei}

\bibitem[{{Merritt}(2013{\natexlab{b}})}]{merritt2013CQG}
{Merritt} D., 2013{\natexlab{b}}, Classical and Quantum Gravity, 30, 244005

\bibitem[{{Merritt} \&
  {Ferrarese}(2001{\natexlab{a}})}]{merrittferrarese2001sphinfl}
{Merritt} D., {Ferrarese} L., 2001{\natexlab{a}}, in Astronomical Society of
  the Pacific Conference Series, Vol. 249, The Central Kiloparsec of Starbursts
  and AGN: The La Palma Connection, {Knapen} J.~H., {Beckman} J.~E., {Shlosman}
  I., {Mahoney} T.~J., eds., p. 335

\bibitem[{{Merritt} \&
  {Ferrarese}(2001{\natexlab{b}})}]{merrittferrarese2001ms}
{Merritt} D., {Ferrarese} L., 2001{\natexlab{b}}, \apj, 547, 140

\bibitem[{{Merritt} {et~al}\mbox{.}(2004){Merritt}, {Milosavljevi{\'c}},
  {Favata}, {Hughes}, \& {Holz}}]{merritt2004}
{Merritt} D., {Milosavljevi{\'c}} M., {Favata} M., {Hughes} S.~A., {Holz}
  D.~E., 2004, \apjl, 607, L9

\bibitem[{{Milosavljevi{\'c}} \& {Merritt}(2001)}]{milosavljevicmerritt2001}
{Milosavljevi{\'c}} M., {Merritt} D., 2001, \apj, 563, 34

\bibitem[{{Moellenhoff}, {Matthias} \& {Gerhard}(1995){Moellenhoff},
  {Matthias}, \& {Gerhard}}]{moellenhoff1995}
{Moellenhoff} C., {Matthias} M., {Gerhard} O.~E., 1995, \aap, 301, 359

\bibitem[{{Morrison} {et~al}\mbox{.}(2011){Morrison}, {Caldwell}, {Schiavon},
  {Athanassoula}, {Romanowsky}, \& {Harding}}]{morrison2011m31}
{Morrison} H., {Caldwell} N., {Schiavon} R.~P., {Athanassoula} E., {Romanowsky}
  A.~J., {Harding} P., 2011, \apjl, 726, L9

\bibitem[{{Naab}, {Johansson} \& {Ostriker}(2009){Naab}, {Johansson}, \&
  {Ostriker}}]{naab2009}
{Naab} T., {Johansson} P.~H., {Ostriker} J.~P., 2009, \apjl, 699, L178

\bibitem[{{Nipoti}, {Londrillo} \& {Ciotti}(2003){Nipoti}, {Londrillo}, \&
  {Ciotti}}]{nipoti2003}
{Nipoti} C., {Londrillo} P., {Ciotti} L., 2003, \mnras, 342, 501

\bibitem[{{Peng}(2007)}]{peng2007}
{Peng} C.~Y., 2007, \apj, 671, 1098

\bibitem[{{Redmount} \& {Rees}(1989)}]{redmountrees1989}
{Redmount} I.~H., {Rees} M.~J., 1989, Comments on Astrophysics, 14, 165

\bibitem[{{Rodriguez} {et~al}\mbox{.}(2006){Rodriguez}, {Taylor}, {Zavala},
  {Peck}, {Pollack}, \& {Romani}}]{rodriguez2006}
{Rodriguez} C., {Taylor} G.~B., {Zavala} R.~T., {Peck} A.~B., {Pollack} L.~K.,
  {Romani} R.~W., 2006, \apj, 646, 49

\bibitem[{{Rusli} {et~al}\mbox{.}(2013{\natexlab{a}}){Rusli}, {Erwin},
  {Saglia}, {Thomas}, {Fabricius}, {Bender}, \& {Nowak}}]{rusli2013}
{Rusli} S.~P., {Erwin} P., {Saglia} R.~P., {Thomas} J., {Fabricius} M.,
  {Bender} R., {Nowak} N., 2013{\natexlab{a}}, \aj, 146, 160

\bibitem[{{Rusli} {et~al}\mbox{.}(2013{\natexlab{b}}){Rusli}, {Thomas},
  {Saglia}, {Fabricius}, {Erwin}, {Bender}, {Nowak}, {Lee}, {Riffeser}, \&
  {Sharp}}]{rusli2013bhmassesDM}
{Rusli} S.~P. {et~al.}, 2013{\natexlab{b}}, \aj, 146, 45

\bibitem[{{Sani} {et~al}\mbox{.}(2011){Sani}, {Marconi}, {Hunt}, \&
  {Risaliti}}]{sani2011}
{Sani} E., {Marconi} A., {Hunt} L.~K., {Risaliti} G., 2011, \mnras, 413, 1479

\bibitem[{{Schechter}(1980)}]{schechter1980}
{Schechter} P.~L., 1980, \aj, 85, 801

\bibitem[{{Scott} {et~al}\mbox{.}(2014){Scott}, {Davies}, {Houghton},
  {Cappellari}, {Graham}, \& {Pimbblet}}]{scott2014}
{Scott} N., {Davies} R.~L., {Houghton} R.~C.~W., {Cappellari} M., {Graham}
  A.~W., {Pimbblet} K.~A., 2014, \mnras, 441, 274

\bibitem[{{Scott}, {Graham} \& {Schombert}(2013){Scott}, {Graham}, \&
  {Schombert}}]{scott2013}
{Scott} N., {Graham} A.~W., {Schombert} J., 2013, \apj, 768, 76

\bibitem[{{Sillanpaa} {et~al}\mbox{.}(1988){Sillanpaa}, {Haarala}, {Valtonen},
  {Sundelius}, \& {Byrd}}]{sillanpaa1988}
{Sillanpaa} A., {Haarala} S., {Valtonen} M.~J., {Sundelius} B., {Byrd} G.~G.,
  1988, \apj, 325, 628

\bibitem[{{Thomas} {et~al}\mbox{.}(2014){Thomas}, {Saglia}, {Bender}, {Erwin},
  \& {Fabricius}}]{thomas2014}
{Thomas} J., {Saglia} R.~P., {Bender} R., {Erwin} P., {Fabricius} M., 2014,
  \apj, 782, 39

\bibitem[{{Trujillo} {et~al}\mbox{.}(2004){Trujillo}, {Erwin}, {Asensio Ramos},
  \& {Graham}}]{trujillo2004coresersicmodel}
{Trujillo} I., {Erwin} P., {Asensio Ramos} A., {Graham} A.~W., 2004, \aj, 127,
  1917

\bibitem[{{Valluri}, {Merritt} \& {Emsellem}(2004){Valluri}, {Merritt}, \&
  {Emsellem}}]{valluri2004}
{Valluri} M., {Merritt} D., {Emsellem} E., 2004, \apj, 602, 66

\bibitem[{{van den Bosch} {et~al}\mbox{.}(2012){van den Bosch}, {Gebhardt},
  {G{\"u}ltekin}, {van de Ven}, {van der Wel}, \& {Walsh}}]{vandenbosch2012}
{van den Bosch} R.~C.~E., {Gebhardt} K., {G{\"u}ltekin} K., {van de Ven} G.,
  {van der Wel} A., {Walsh} J.~L., 2012, \nat, 491, 729

\bibitem[{{Volonteri} \& {Ciotti}(2013)}]{volontericiotti2013}
{Volonteri} M., {Ciotti} L., 2013, \apj, 768, 29

\bibitem[{{von der Linden} {et~al}\mbox{.}(2007){von der Linden}, {Best},
  {Kauffmann}, \& {White}}]{vonderlinden2007}
{von der Linden} A., {Best} P.~N., {Kauffmann} G., {White} S.~D.~M., 2007,
  \mnras, 379, 867

\end{thebibliography}

\label{lastpage}

\end{document}